\documentclass[aps,twocolumn,showkeys,showpacs,preprintnumbers,prd,superscriptaddress,nofootinbib,10pt]{revtex4-1}
\usepackage{graphicx,epsf,bm,amsmath,amsfonts,amssymb,epstopdf,natbib,hyperref,xcolor,appendix,verbatim, multirow,diagbox,booktabs,siunitx}
\usepackage[normalem]{ulem}
\usepackage{array}
\usepackage{diagbox}
\usepackage{makecell}
\usepackage[makeroom]{cancel}
\usepackage{fontawesome}
\usepackage{hyperref}
\hypersetup{colorlinks=true,urlcolor=blue,citecolor=blue,linkcolor=blue,menucolor=blue,anchorcolor=blue,filecolor=blue}

\newcolumntype{?}{!{\vrule width 3pt}}

\usepackage{graphicx} % Required for \scalebox
\usepackage{array}    % Required for custom column types
\usepackage{makecell} % Required for \Xhline

\newcolumntype{?}{!{\vrule width 1.5pt}}

\date{\today}

\def\@fnsymbol#1{{\ifcase#1\or \fmarki\or \fmarkii\or \fmarkiii\or \fmarkiv\or \fmarkv\or \else\@ctrerr\fi}}
\makeatother

\begin{document}

\title{When, Why, and How CMB Compression Fails}

\author{Davide Pedrotti}
\email{davide.pedrotti-1@unitn.it}
\affiliation{Department of Physics, University of Trento, Via Sommarive 14, 38123 Povo (TN), Italy}
\affiliation{Trento Institute for Fundamental Physics and Applications (TIFPA)-INFN, Via Sommarive 14, 38123 Povo (TN), Italy}

\author{William Giarè}
\email{giare@hawaii.edu}
\affiliation{Department of Physics and Astronomy, University of Hawai‘i, Honolulu, HI 96822, USA \looseness=-1}

\author{Hanyu Cheng}
\email{chenghanyu@sjtu.edu.cn}
\affiliation{Tsung-Dao Lee Institute \& School of Physics and Astronomy, Shanghai Jiao Tong University, Shanghai 201210, China \looseness=-1}

\author{Eleonora Di Valentino}
\email{e.divalentino@sheffield.ac.uk}
\affiliation{School of Mathematical and Physical Sciences, University of Sheffield, Hounsfield Road, Sheffield S3 7RH, United Kingdom \looseness=-1}

\begin{abstract}
\noindent We test when compressed CMB likelihoods can be reliably used to constrain new physics beyond $\Lambda$CDM, formulating concrete criteria for \textit{when}, \textit{why}, and \textit{how} CMB compression fails. We derive our conclusions by exploring a broad set of models that modify different aspects of the cosmological evolution, including late-time dark-energy dynamics, non-standard late-time dark-matter evolution through interactions or decay, and dark-matter models with persistently non-standard redshift evolution. For each case, we compare the results obtained from data combinations based on the full Planck-CMB likelihood with their counterparts based on a compressed CMB likelihood calibrated in $\Lambda$CDM. We find that the compression remains accurate for late-time dark-energy extensions that preserve the dark-matter background evolution and do not introduce significant additional perturbative effects. By contrast, modifying the dark-matter evolution breaks the mapping between the present-day matter density and the matter density at the epochs relevant to the CMB, causing the compressed constraints to fail even when the modification acts exclusively at late times. In these cases, the compressed likelihood can drastically alter the geometry of the multidimensional posterior distributions, shifting their centers, producing strongly anisotropic deformations, reordering principal modes, and generating genuine mixing and rotation of their eigendirections. These multidimensional distortions propagate into substantially incorrect marginalized constraints on cosmological parameters, with differences between the full and compressed analyses reaching nearly an order of magnitude in some of the cases studied here.
\end{abstract}

\maketitle

%===================
\section{Introduction}
\label{sec:intro}

Over more than two decades, full-sky satellite missions such as WMAP~\cite{WMAP:2012fli,WMAP:2012nax} and Planck~\cite{Planck:2018nkj,Planck:2018vyg}, complemented by the higher angular resolution of ground-based experiments such as the Atacama Cosmology Telescope (ACT)~\cite{ACT:2023dou,ACT:2023kun,AtacamaCosmologyTelescope:2025blo} and the South Pole Telescope (SPT)~\cite{SPT-3G:2024atg,SPT-3G:2025bzu}, have established the Cosmic Microwave Background (CMB) radiation as one of the most powerful and precise probes in modern cosmology.

The increasing precision of CMB measurements has progressively tightened constraints on cosmological parameters and models of new physics, while confirming the excellent agreement of the observed anisotropies with the standard $\Lambda$CDM cosmology~\cite{Planck:2018vyg,ACT:2023kun,SPT-3G:2025bzu}. At the same time, however, persistent tensions within $\Lambda$CDM, including the Hubble tension~\cite{Verde:2019ivm,DiValentino:2020zio,DiValentino:2021izs,Perivolaropoulos:2021jda,Schoneberg:2021qvd,Shah:2021onj,Abdalla:2022yfr,DiValentino:2022fjm,Kamionkowski:2022pkx,Giare:2023xoc,Hu:2023jqc,Verde:2023lmm,DiValentino:2024yew,Efstathiou:2024dvn,CosmoVerseNetwork:2025alb,Cai:2026swf} and the preference for dynamical dark energy recently emerging from DESI analyses~\cite{DESI:2025zgx,DESI:2025fii},\footnote{For a non-exhaustive selection of discussions surrounding this preference, see, e.g.~\cite{Cortes:2024lgw,Shlivko:2024llw,Luongo:2024fww,Gialamas:2024lyw,Wang:2024dka,Ye:2024ywg,Tada:2024znt,Carloni:2024zpl,Chan-GyungPark:2024mlx,Efstathiou:2024xcq,Bhattacharya:2024hep,Andriot:2024jsh,Li:2024qus,Reboucas:2024smm,Giare:2024gpk,Ramadan:2024kmn,Yang:2024kdo,Jiang:2024xnu,RoyChoudhury:2024wri,Giare:2024oil,Giare:2025pzu,Efstathiou:2025tie,Kessler:2025kju,RoyChoudhury:2025dhe,Scherer:2025esj,Wolf:2025jlc,Santos:2025wiv,Specogna:2025guo,Cheng:2025lod,Cheng:2025hug,Ozulker:2025ehg,Li:2025vuh,Lee:2025pzo,Fazzari:2025lzd,Teixeira:2025czm,Smith:2025icl,Cai:2025mas,Herold:2025hkb,Cheng:2025yue,Chudaykin:2025aux,Chudaykin:2025lww,Chudaykin:2025vdh,Ivanov:2026dvl,Chudaykin:2026nls,Gokcen:2026pkq,Ishak:2025cay,Najafi:2026kxs,Yang:2026yaq,Dinda:2026ktu,Dainotti:2026enb,Kessler:2026dbi,Li:2026xaz,Giare:2026tyk,Li:2026asg,Ren:2026jyw,Andriot:2026lac}.} have motivated growing interest in alternative models and physical mechanisms beyond the standard cosmological paradigm.\footnote{The significance of the so-called $S_8$ tension~\cite{DiValentino:2020vvd,DiValentino:2018gcu,Nunes:2021ipq,DES:2021bvc,DES:2021vln,KiDS:2020suj,Asgari:2019fkq,Joudaki:2019pmv,DAmico:2019fhj,Kilo-DegreeSurvey:2023gfr,Troster:2019ean,Heymans:2020gsg,Dalal:2023olq,Chen:2024vvk,ACT:2024okh,DES:2024oud,Harnois-Deraps:2024ucb,Dvornik:2022xap,DES:2021wwk,Wright:2025xka,DES:2026fyc} is milder and less clear-cut. The KiDS-Legacy reanalysis reported values of $S_8$ in good agreement with Planck~\cite{Wright:2025xka}, whereas the Dark Energy Survey Year 6 (DES-Y6) results still find a mild mismatch with CMB constraints at $\sim 2.6$--$2.8\sigma$~\cite{DES:2026fyc}.} Since any such mechanism must preserve the level of agreement with CMB data set by $\Lambda$CDM, confronting models of new physics with CMB observations is more important and timely than ever.

However indispensable in practice, testing models of new physics against CMB data has two important limitations. First, it is computationally expensive and inconvenient for efficient scans across large model spaces. Second, it is not easily applicable to models for which the evolution of linear perturbations is either technically and computationally demanding (as often occurs in modified-gravity theories~\cite{Clifton:2011jh,Koyama:2015vza}) or not well defined (as in some purely phenomenological models).

These limitations have motivated the development of \textit{compression schemes}, which summarize the information contained in the full CMB power spectrum in a limited number of parameters~\cite{Efstathiou:1998xx}. The most notable example is given by the so-called shift parameters ($\ell_a$, $R$), which were first introduced with the aim of exploring cosmic acceleration and the landscape of dark energy models~\cite{Efstathiou:1998xx,Wang:2006ts,Wang:2007mza}. $R\equiv \sqrt{\Omega_mH_0^2}D_M(z_\star)$ represents the \textit{scaled} angular diameter distance to the last-scattering surface, while $\ell_a \equiv \pi D_M(z_\star)/r_s^\star$ defines the multipole corresponding to the characteristic angular scale subtended by the sound horizon at photon decoupling. In Refs.~\cite{Wang:2007mza,Wang:2013mha} it was subsequently shown that combining the shift parameters with the physical baryon density $\omega_b = \Omega_b h^2$ (and possibly the inflationary spectral index $n_s$) provides a compact summary of the CMB information relevant for cosmological parameter constraints. This set of quantities is commonly referred to as the CMB \textit{distance priors}.\footnote{Other methods to compress CMB information have also been explored. For example, in Ref.~\cite{Prince:2019hse} the authors applied the Massively Optimized Parameter Estimation and Data compression to the \textit{Planck-2015} temperature likelihood and showed that the standard $\Lambda$CDM cosmological parameter constraints can be derived from a compressed likelihood of just six Gaussian-distributed data points. More recently, Ref.~\cite{Akhmetzhanova:2023hiy} explored self-supervised machine-learning techniques for data compression and cosmological inference.}

Despite being among the least model-dependent quantities that can be extracted from the CMB, the shift parameters are not direct observables. Quantities such as $\ell_a$ and $R$ are inferred from the observed temperature and polarization power spectra within a fiducial cosmological model, typically $\Lambda$CDM~\cite{Wang:2007mza,Mukherjee:2008kd,Wang:2013mha}. Their values and correlations therefore depend on how the fiducial model relates cosmological parameters to the observed CMB features~\cite{Zhai:2019nad}. Using the resulting compressed likelihood to constrain a different cosmology implicitly assumes that the relevant CMB physics, and hence this mapping, remains sufficiently unchanged~\cite{Mukherjee:2008kd,Zhai:2019nad}. A commonly adopted working criterion is to distinguish between modifications occurring before and after recombination. Late-time models are generally expected to be more reliable in this respect, since they leave the physics responsible for the primary CMB anisotropies essentially unchanged~\cite{Wang:2007mza,Mukherjee:2008kd,Zhai:2019nad}. By contrast, pre-recombination new physics can modify the acoustic physics itself and therefore invalidate a compression calibrated within $\Lambda$CDM~\cite{Shi:2015tje,Zhai:2019nad}.

A number of studies have therefore assessed the validity and regime of applicability of compressed CMB information beyond the fiducial $\Lambda$CDM cosmology. Initial studies tested the robustness of the shift parameters against relatively simple departures from flat $\Lambda$CDM~\cite{Wang:2007mza,Mukherjee:2008kd,Wang:2013mha}, while later analyses extended these tests to models involving dynamical dark energy, massive neutrinos, additional relativistic degrees of freedom, spatial curvature, and modifications of the primordial power spectrum~\cite{Zhai:2019nad}. More recently, the compressed likelihood \texttt{CMBComp}~\cite{Srivastav:2026toe} has addressed this model dependence more systematically by constructing compression schemes tailored to specific classes of cosmological models.\footnote{For instance, the $\nu\Lambda$CDM model is described by the four-parameter compression $(\ell_a,R,\omega_b,\omega_c)$, calibrated specifically within the corresponding model space.}

Our work extends these robustness tests with the broader goal of identifying the physical conditions under which compressed CMB information can be reliably used to test models of new physics. In particular, we investigate which properties of a cosmological model determine whether compressed CMB information remains reliable, focusing on the physical sector affected by the new dynamics and on which pieces of CMB information are consequently modified.

To this end, we consider a set of phenomenologically motivated models proposed in connection with concrete challenges to the standard cosmological paradigm and use them as representative test cases to isolate qualitatively different departures from the physics underlying the CMB compression. Taking $\Lambda$CDM as our benchmark scenario, we first extend it by introducing a constant dark-energy equation of state (EoS) $w_{\rm DE}$. We then consider the Chevallier-Polarski-Linder (CPL) parametrization of dynamical dark energy~\cite{Chevallier:2000qy,Linder:2002et} and a sign-switching cosmological constant~\cite{Akarsu:2021fol}. These three models modify the late-time dark-energy background while leaving the standard scaling of the matter densities unchanged. We then turn to interacting dark-energy (IDE) models, where energy exchange within the dark sector alters the dynamics of both dark energy and dark matter~\cite{Bolotin:2013jpa,Wang:2016lxa}, and consequently the relation between the dark-matter abundance constrained by the CMB at recombination and its present-day value. Finally, we consider two complementary scenarios in which the departure from $\Lambda$CDM is confined to the dark-matter sector: decaying dark matter~\cite{Audren:2014bca,Poulin:2016nat} and dark matter with a nonzero EoS~\cite{Kopp:2018zxp}. For the latter, we consider both a time-dependent EoS, $w_{\rm dm}(a)\propto a$, and a constant EoS, $w_{\rm dm}(a)=w_{\rm dm,0}$. While dark-matter decay and the time-dependent EoS affect the dark-matter evolution primarily at late times, a constant nonzero EoS modifies its redshift scaling throughout cosmic history. Together, these models allow us to test whether the validity of the CMB compression depends not simply on when the new physics becomes relevant, but on how it modifies the relation between the dark-matter abundance probed by the CMB and its present-day value.

We find a clear qualitative distinction depending on the physical sector affected by the new dynamics. The compression remains remarkably robust for models in which the departure from $\Lambda$CDM is confined to the late-time dark-energy sector and the standard evolution of the matter densities is preserved. By contrast, when the dark-matter sector acquires a non-standard evolution, the compressed likelihood fails to reproduce the constraints obtained from the full CMB temperature and polarization spectra. In these cases, the compressed likelihood can distort the multidimensional geometry of the posterior distributions, ultimately leading to marginalized constraints that in some cases differ by nearly an order of magnitude from those obtained with the full CMB likelihood. We trace this breakdown primarily to the failure of the usual mapping between the dark-matter abundance constrained by the CMB around recombination and its present-day value. In addition, compressing the full spectra into a small number of background quantities can discard perturbation-level information that becomes relevant when the dark-matter sector is modified. Our results therefore show that the validity of compressed CMB information is determined not simply by \textit{when} new physics becomes active, but also by \textit{which physical sector it affects} and whether the relations implicitly assumed by the compression remain valid.

The paper is organized as follows. In Sec.~\ref{sec:compression} we review the compression scheme adopted in our analysis. In Sec.~\ref{sec:methodology} we discuss the datasets and methodology used to obtain our results, together with the diagnostic tools developed to perform a detailed comparison of the posterior distributions. In Sec.~\ref{sec:models} we introduce the models against which we test the CMB compression. We present the results in Sec.~\ref{sec:results}. A critical discussion of our findings and their implications follows in Sec.~\ref{sec:discussion}. Finally, we present our conclusions in Sec.~\ref{sec:conclusions}. Additional tests of the robustness of our main conclusions against different compression choices and dataset combinations are presented in Appendix~\ref{Appendix:A}.
%===================

\section{CMB compression}
\label{sec:compression}

In this section, we give a brief introduction to the shift parameters $(R,\ell_a)$, which together with the physical baryon density $\omega_b$ constitute the CMB distance priors. We then describe their relation to the compression scheme adopted in this paper.

The first shift parameter, $R$, is defined as~\cite{Bond:1997wr}
\begin{equation}
R \equiv \sqrt{\Omega_m H_0^2}\,D_M(z_\star),
\label{eq:R}
\end{equation}
where $H_0$ is the Hubble constant, $\Omega_m$ is the present-day matter density parameter, and $D_M(z_\star)$ is the comoving distance to the last-scattering surface. Throughout this work, we consider a spatially flat, homogeneous and isotropic Universe described by a Friedmann-Lemaître-Robertson-Walker metric. In this framework, the comoving distance to the last-scattering surface reads
\begin{equation}
D_M(z_\star)=\frac{1}{H_0}\int_0^{z_\star}\frac{dz'}{E(z')},
\end{equation}
where $E(z)\equiv H(z)/H_0$ is the dimensionless expansion rate of the Universe, and we work in units where $c=1$. The redshift $z_\star$ denotes photon decoupling and is conventionally identified with the peak of the Thomson-scattering visibility function. Both $z_\star$ and $D_M(z_\star)$ depend on the assumed cosmology and can be computed numerically using Boltzmann solvers such as CAMB~\cite{Lewis:1999bs} or CLASS~\cite{Lesgourgues:2011re,Blas:2011rf}. Historically, the shift parameter $R$ was introduced to capture the dominant effect of late-time geometry on the angular structure of the CMB power spectrum. Ref.~\cite{Bond:1997wr} showed that models with the same primordial perturbation spectrum and the same physical matter densities at decoupling can produce nearly identical primary CMB spectra when their angular-distance scaling is also preserved.

The second shift parameter, $\ell_a$, is defined as~\cite{Elgaroy:2007bv}
\begin{equation}
\ell_a \equiv \frac{\pi}{\theta_s^\star},
\end{equation}
where $\theta_s^\star \equiv r_s^\star /D_M(z_\star)$ is the angular acoustic scale subtended by the comoving sound horizon at photon decoupling, $r_s^\star$. Interestingly, $\theta_s^\star$ is determined by the ratio of two quantities sensitive to different epochs. The comoving distance $D_M(z_\star)$ depends on the integrated expansion history between the last-scattering surface and today. By contrast, the comoving sound horizon at photon decoupling is determined by the pre-recombination expansion history,
\begin{equation}
r_s^\star = \int_{z_\star}^{\infty} dz'\,\frac{c_s(z')}{H(z')},
\end{equation}
where $c_s(z)$ denotes the sound speed of the tightly coupled photon-baryon plasma given by
\begin{equation}
c_s(z)=\frac{1}{\sqrt{3\left[1+R_b(z)\right]}},
\quad
R_b(z)\equiv
\frac{3\omega_b}{4\omega_\gamma}\frac{1}{1+z}.
\end{equation}
Although both $r_s^\star$ and $D_M(z_\star)$ are model dependent, their ratio is exceptionally tightly constrained by the observed acoustic pattern of the CMB. As a result, $\theta_s^\star$ remains remarkably stable across broad classes of models that preserve the underlying acoustic physics, making $\ell_a$ particularly well suited for CMB compression.

The shift parameters $R$ and $\ell_a$ capture the dominant geometrical information encoded in the CMB, but an accurate compression must also retain the physical quantities that determine the acoustic structure at recombination. This role is naturally played by the physical baryon density $\omega_b$. Its value directly affects the sound speed of the photon-baryon fluid, the sound horizon, and the relative heights and structure of the acoustic peaks in the CMB spectra. Together, these effects tightly constrain $\omega_b$ and generate correlations with the shift parameters that must be preserved in the compression scheme. This leads to the three-parameter distance-prior set $(R,\ell_a,\omega_b)$ considered in Ref.~\cite{Wang:2007mza}.\footnote{Additional parameters may be required depending on the information included in the analysis. A notable example is the scalar spectral index, $n_s$, whose inclusion becomes relevant when using observables sensitive to the shape of the matter power spectrum~\cite{Mukherjee:2008kd}. Since in our analysis BAO measurements are used only as geometrical probes, we do not retain $n_s$ as an additional compression parameter~\cite{Mukherjee:2008kd}.}

In our work, we adopt the compression vector $(1/\theta_s^{\star},\omega_m,\omega_b)$, where $\omega_m=\omega_c+\omega_b+\omega_\nu$. Our compression vector is closely related to the standard distance-prior set $(R,\ell_a,\omega_b)$.\footnote{Similar parametrizations have been successfully adopted in Refs.~\cite{Lin:2021sfs,Lynch:2025ine,Pedrotti:2025ccw,Weiner:2026sfm}.} For cosmologies that preserve the standard pre-recombination physics, the two parametrizations are expected to encode equivalent information. For our purposes, however, the adopted parametrization offers a more transparent physical interpretation by separating the geometrical information carried by the angular acoustic scale from that associated with the physical matter densities. Under standard pre-recombination physics and for fixed radiation and neutrino sectors, the sound horizon $r_s^\star$ is determined by the physical densities $\omega_m$ and $\omega_b$. Once this early-time calibration is specified, $1/\theta_s^{\star}=D_M(z_\star)/r_s^\star$ isolates the remaining dependence on the distance to the last-scattering surface and hence on the post-recombination expansion history. The vector $(1/\theta_s^{\star},\omega_m,\omega_b)$ therefore provides a clean separation between the geometrical information contained in $D_M(z_\star)$ and the physical densities that determine the acoustic ruler against which this distance is measured. This separation is less explicit in the traditional basis $(R,\ell_a,\omega_b)$. Indeed, $R$ combines $D_M(z_\star)$ with the present-day matter density, while $\ell_a$ also depends on $D_M(z_\star)$ through the angular acoustic scale. The two parameters therefore share the same dependence on the distance to last scattering. Replacing $R$ with $\omega_m$ removes this common geometrical dependence and makes the separation between geometry and physical densities explicit. This separation becomes particularly informative when the evolution of the matter sector itself is modified. For completeness, in Appendix~\ref{sec:WangWang} we explicitly verify the equivalence between the two formulations of the CMB distance priors.
%===================

\section{dataset and methodology}
\label{sec:methodology}

In this section, we describe the datasets and methodology adopted in our analysis. We first summarize the numerical implementation and statistical framework used for cosmological parameter inference. We then describe the tailored geometric posterior analysis developed to quantify the differences between the full and compressed CMB constraints, together with its numerical implementation.

\subsection{Datasets and standard parameter inference}
\label{sec:datasets}

We perform a systematic comparison of cosmological constraints obtained using the full and compressed CMB likelihoods across a set of representative models chosen to isolate qualitatively different departures from $\Lambda$CDM and identify the physical properties that determine the validity of the CMB compression. Theoretical predictions are computed using the Boltzmann solver \texttt{CLASS}~\cite{Blas:2011rf,Lesgourgues:2011re}. We use the standard public implementation for $\Lambda$CDM and for scenarios already supported by the code, together with dedicated modified versions for the non-standard dark-energy and dark-matter models introduced in Sec.~\ref{sec:models}. These implementations consistently evolve the background cosmology and, whenever required for the full-CMB analysis, the associated linear perturbations.

For each model, we run parallel Markov Chain Monte Carlo (MCMC) analyses using \texttt{Cobaya}~\cite{Torrado:2020dgo}. The chains are evolved until convergence according to the Gelman-Rubin criterion~\cite{Gelman:1992zz}, requiring $R-1<0.01$, and subsequently post-processed using \texttt{GetDist}~\cite{Lewis:2019xzd} to derive marginalized parameter constraints and posterior distributions. We further analyze the multidimensional geometry of the posterior distributions using the \texttt{PostModes} package described in the following subsection.

The full- and compressed-CMB analyses use the same cosmological parameters, prior distributions, and observational datasets. They differ only in the treatment of the CMB information. In one case, we use the full Planck temperature and polarization likelihood, while in the other we replace it with the compressed likelihood introduced in Sec.~\ref{sec:compression}. The observational datasets and likelihoods entering these analyses are the following:

\begin{itemize}
\item \textbf{Full Planck-CMB}: Our full CMB likelihood is based on the Planck 2018 data~\cite{Planck:2019nip,Planck:2018vyg} and combines the low-$\ell$ temperature \texttt{Commander} likelihood, the low-$\ell$ E-mode polarization \texttt{SimAll} likelihood, and the high-$\ell$ \texttt{Plik} TT, TE, and EE likelihood~\cite{Planck:2019nip}. We do not include the CMB lensing-reconstruction likelihood because our goal is to compare the information encoded in the primary CMB spectra with that retained by their compression. Including lensing reconstruction would add extra late-time information on structure growth that is not encoded in the compressed likelihood. Gravitational-lensing effects on the primary TT, TE, and EE spectra are nevertheless consistently included.

\item \textbf{Compressed Planck-CMB}: We perform an MCMC analysis of $\Lambda$CDM using exclusively the Planck 2018 temperature and polarization data listed above. From the resulting MCMC chains, following the parametrization introduced in Sec.~\ref{sec:compression}, we extract the posterior mean $\boldsymbol{\mu}$ and covariance $\mathbf{C}$ of the three-dimensional compression vector $\mathbf{x}=(1/\theta_s,\omega_m,\omega_b)$. We then use these quantities to construct a compressed Planck multivariate Gaussian likelihood,\footnote{Available at \faGithub\ \url{https://github.com/williamgiare/wgcosmo}.}
\begin{equation}
-2\ln\mathcal{L}_{\rm CMB}^{3\times3}
=
\left(\mathbf{x}-\boldsymbol{\mu}\right)^{\rm T}
\mathbf{C}^{-1}
\left(\mathbf{x}-\boldsymbol{\mu}\right),
\label{eq:compressed_likelihood}
\end{equation}
which preserves the means, variances, and correlations of these three quantities.

\item \textbf{DESI}: We use the Dark Energy Spectroscopic Instrument (DESI) DR2 Baryon Acoustic Oscillation (BAO) measurements from galaxies, quasars, and Ly$\alpha$-forest data~\cite{DESI:2025qqy,DESI:2025zpo} in seven redshift bins spanning $0.295\leq z_{\rm eff}\leq 2.330$~\cite{DESI:2025zgx}. The likelihood includes the full covariance matrix of the measurements~\cite{DESI:2025zgx}.

\item \textbf{PantheonPlus}: We use Type-Ia Supernovae (SN) distance-modulus measurements from the \textit{Pantheon+} sample, consisting of 1701 light curves from 1550 unique spectroscopically classified SN~\cite{Brout:2022vxf}.
\end{itemize}

Throughout this work, we denote by
\begin{itemize}
    \item $A$: the posterior and marginalized constraints obtained from the full Planck-CMB likelihood combined with DESI BAO and PantheonPlus SN;
    \item $B$: the posterior and marginalized constraints obtained from the compressed Planck-CMB likelihood combined with DESI BAO and PantheonPlus SN.
\end{itemize}

\subsection{Posterior eigenmode analysis}
To characterize more systematically the differences between the full and compressed CMB constraints, we complement the standard comparison of marginalized constraints on cosmological parameters with a broader geometric analysis of the corresponding multidimensional posteriors.

To this end we develop the \texttt{PostModes} package.\footnote{Available at \faGithub\ \url{https://github.com/williamgiare/postmodes}.} It compares two posterior distributions, $p_A(\boldsymbol{\theta})$ and $p_B(\boldsymbol{\theta})$, defined over the same set of $N$ parameters. Starting from their posterior means and covariance matrices, it quantifies the differences between $B$ and $A$ by characterizing shifts in the posterior centers, changes in the orientation of the principal degeneracy directions, and variations in the relative constraining power along specific parameter combinations. It also separates changes in the overall scale of the posterior covariance from direction-dependent deformations. The method requires finite second moments and positive-definite covariance matrices, but does not assume Gaussian posterior distributions.\footnote{For non-Gaussian posteriors, the resulting diagnostics characterize their second-moment geometry rather than their complete shapes or highest-posterior-density regions.}

In the concrete application considered here, $A$ denotes the posterior obtained using the full CMB likelihood, while $B$ denotes the corresponding posterior obtained using the compressed CMB likelihood. The comparison is based on the posterior mean vectors $\boldsymbol{\mu}_A$, $\boldsymbol{\mu}_B$ and covariance matrices $\mathbf{C}_A$, $\mathbf{C}_B$, extracted directly from the corresponding MCMC chains. In the following, we define the diagnostic quantities used to characterize their agreement and discuss what each of them reveals about the impact of CMB compression on the multidimensional posterior. Additional documentation, worked examples, and validation tests are provided with the \texttt{PostModes} package.

\subsubsection{Individual posterior geometry}
The covariance geometry of the two posteriors is characterized through the principal-component decompositions
\begin{equation}
\mathbf{C}_A=\mathbf{E}_A\mathbf{\Lambda}_A\mathbf{E}_A^T,
\qquad
\mathbf{C}_B=\mathbf{E}_B\mathbf{\Lambda}_B\mathbf{E}_B^T,
\end{equation}
where $\mathbf{E}_A$ and $\mathbf{E}_B$ define the principal directions of the corresponding posterior covariances, while the eigenvalues contained in $\mathbf{\Lambda}_A$ and $\mathbf{\Lambda}_B$ determine the variances along those directions. This decomposition provides a first characterization of the parameter combinations that are most and least constrained by the full- and compressed-CMB posterior distributions. Since the resulting eigenvectors depend on the adopted parametrization and units, we do not interpret their individual coefficients as coordinate-independent measures of physical importance.

\subsubsection{Displacement of the posterior centers}
We define the displacement between the posterior means $\Delta\boldsymbol{\mu}=\boldsymbol{\mu}_B-\boldsymbol{\mu}_A$
and quantify it through the Mahalanobis distances
\begin{align}
&S_{B|A}
=
\left[
\Delta\boldsymbol{\mu}^T
\mathbf{C}_A^{-1}
\Delta\boldsymbol{\mu}
\right]^{1/2},
\\
&S_{A|B}
=
\left[
\Delta\boldsymbol{\mu}^T
\mathbf{C}_B^{-1}
\Delta\boldsymbol{\mu}
\right]^{1/2},
\\
&S_{\rm{CC}}
=
\left[
\Delta\boldsymbol{\mu}^T
(\mathbf{C}_A+\mathbf{C}_B)^{-1}
\Delta\boldsymbol{\mu}
\right]^{1/2}.
\end{align}
The quantities $S_{B|A}$ and $S_{A|B}$ measure the same displacement using the covariance of posterior $A$ or $B$ as the corresponding metric, respectively. The symmetric combined-covariance distance (SCCD), $S_{\rm CC}$, instead provides an exchange-symmetric measure of the separation between the posterior centers. For independent Gaussian posteriors, the squared SCCD can be interpreted as a $\chi^2$ statistic and converted into a Gaussian-equivalent significance.\footnote{For an $N$-dimensional comparison, $S_{\rm CC}^2\sim\chi_N^2$, so that $p=1-F_{\chi_N^2}(S_{\rm CC}^2)$ and the corresponding two-sided Gaussian-equivalent significance is $\sigma_{\rm eq}=\Phi^{-1}(1-p/2)$, where $F_{\chi_N^2}$ and $\Phi$ denote the $\chi^2$ and standard-normal cumulative distribution functions, respectively.} In our application, however, the full and compressed analyses are derived from the same underlying observations and are therefore statistically correlated. A calibrated significance would thus require the cross-covariance between the two analyses, so we use the SCCD only as a geometric measure of posterior displacement.\looseness=-1

\subsubsection{Orientation of the posterior degeneracies: mode mixing and rotations}
To compare the orientation of the principal degeneracy directions without introducing an arbitrary dependence on parameter units, we express both covariance matrices in a common reference-scaled parameter space. Defining
\begin{equation}
\mathbf{D}_A=
\operatorname{diag}
\left(
\sqrt{(\mathbf{C}_A)_{11}},
\ldots,
\sqrt{(\mathbf{C}_A)_{NN}}
\right),
\end{equation}
we construct
\begin{equation}
\widehat{\mathbf{C}}_A=
\mathbf{D}_A^{-1}\mathbf{C}_A\mathbf{D}_A^{-1},
\qquad
\widehat{\mathbf{C}}_B=
\mathbf{D}_A^{-1}\mathbf{C}_B\mathbf{D}_A^{-1}.
\end{equation}
The same transformation, defined by the marginal standard deviations of the reference posterior $A$, is applied to both distributions, so that their relative widths and correlations are preserved. Let $\widehat{\boldsymbol{e}}_{A,i}$ and $\widehat{\boldsymbol{e}}_{B,j}$ denote the eigenvectors of $\widehat{\mathbf{C}}_A$ and $\widehat{\mathbf{C}}_B$, respectively. We quantify their relative orientation through the overlap matrix
\begin{equation}
O_{ij}
=
\left|
\widehat{\boldsymbol{e}}_{A,i}^{T}
\widehat{\boldsymbol{e}}_{B,j}
\right|.
\label{eq:overlap}
\end{equation}
Values close to unity indicate aligned principal directions, while significant off-diagonal elements can reveal a reordering or mixing of the posterior degeneracy directions. After accounting for possible mode reordering, reduced overlaps between matched modes indicate a genuine change in their orientation.\footnote{When eigenvalues are nearly degenerate, small perturbations can induce large rotations among the associated eigenvectors, so the comparison should be performed between the subspaces they span rather than individual directions.}

\subsubsection{Relative constraining power: generalized modes}
To identify the parameter combinations whose uncertainties change most strongly when the full CMB likelihood is replaced by its compressed counterpart, we whiten the covariance of posterior $B$ with respect to the covariance of posterior $A$. We define
\begin{equation}
\mathbf{\tilde C}
=
\mathbf{C}_A^{-1/2}\,
\mathbf{C}_B\,
\mathbf{C}_A^{-1/2}.
\end{equation}
This transformation maps the covariance of the full-CMB posterior $A$ to the identity, corresponding to a unit hypersphere in which every direction has unit variance. The matrix $\widetilde{\mathbf{C}}$ then represents the covariance of the compressed-CMB posterior $B$ in the same coordinates. Deviations of $\widetilde{\mathbf{C}}$ from the identity therefore directly quantify how the compression changes the posterior widths along different parameter combinations.

The parameter combinations that diagonalize this relative deformation are obtained from the generalized eigenvalue problem $\mathbf{C}_B\boldsymbol{v}_i=\rho_i\mathbf{C}_A\boldsymbol{v}_i$.
Defining the corresponding modes as $m_i=
\boldsymbol{v}_i^T
(\boldsymbol{\theta}-\boldsymbol{\mu}_A)$ the generalized eigenvalues satisfy
\begin{equation}
\rho_i
=
\frac{\operatorname{Var}_B(m_i)}
{\operatorname{Var}_A(m_i)}.
\end{equation}
They therefore provide a direct measure of the relative constraining power of the two analyses along each mode, with the following interpretation:
\begin{itemize}
    \item $\rho_i>1$: the posterior $B$ is broader than the posterior $A$ along mode $m_i$;
    \item $\rho_i<1$: the posterior $B$ is tighter than the posterior $A$ along mode $m_i$;
    \item $\rho_i=1$: the variance along mode $m_i$ is unchanged between $A$ and $B$.
\end{itemize}
The corresponding ratio of standard deviations is $\sqrt{\rho_i}$. We further define
\begin{equation}
\rho_{\min}\equiv \min_i \rho_i,
\qquad
\rho_{\max}\equiv \max_i \rho_i,
\end{equation}
which identify the smallest and largest relative variance ratios across all modes. They therefore quantify the most extreme changes in constraining power induced by the CMB compression.

\subsubsection{Isotropic scaling and anisotropic deformation}
The generalized eigenvalues can be further decomposed into an overall isotropic rescaling of the posterior covariance and a residual anisotropic deformation. 

We quantify the overall isotropic rescaling of the posterior covariance as
\begin{equation}
\alpha
=
\left(
\frac{\det\mathbf{C}_B}
{\det\mathbf{C}_A}
\right)^{1/N}
=
\exp\left(
\frac{1}{N}
\sum_i\ln\rho_i
\right).
\end{equation}
In the absence of anisotropic deformation, one would have $\mathbf{C}_B=\alpha\,\mathbf{C}_A$, so that $\sqrt{\alpha}$ gives the common rescaling of the posterior standard deviations. More generally, $\alpha$ is the geometric mean of the generalized variance ratios $\rho_i$ and captures the overall change in scale independently of direction-dependent distortions. Values $\alpha>1$ correspond to an overall broadening of posterior $B$ relative to $A$, while $\alpha<1$ corresponds to an overall tightening. The case $\alpha=1$ indicates no net isotropic rescaling, although direction-dependent deformations may still be present.

The remaining anisotropic deformation is quantified by
\begin{equation}
A_{\rm aniso}
=
\left[
\frac{1}{N}
\sum_i
\left(
\ln\rho_i-\ln\alpha
\right)^2
\right]^{1/2}.
\end{equation}
This quantity measures the dispersion of the generalized variance ratios around the isotropic rescaling $\alpha$. It vanishes for a purely uniform rescaling, $\mathbf{C}_B=\alpha\,\mathbf{C}_A$, and increases as the relative change in constraining power becomes more direction dependent.

Taken together, these diagnostics provide a complementary characterization of how the compressed posterior differs from the full-CMB reference. The SCCD quantifies shifts in the posterior center, the overlap matrix tracks changes in the principal degeneracy directions, and the generalized eigenvalues $\rho_i$ quantify directional changes in constraining power. The parameters $\alpha$ and $A_{\rm aniso}$ further separate the overall isotropic rescaling of the posterior covariance from its residual anisotropic deformation.
%===================
\section{Models}
\label{sec:models}

In this section, we introduce the cosmological models used to compare the compressed and full CMB likelihoods.

\subsection{$\Lambda$CDM}

The standard spatially flat $\Lambda$CDM model serves as the benchmark cosmology used to calibrate and validate the compressed CMB likelihood. Its background expansion reads
\begin{equation}
E^2(a) = \Omega_r\, a^{-4} + \Omega_m\, a^{-3} + \Omega_\Lambda,
\end{equation}
with $\Omega_m+\Omega_r+\Omega_\Lambda=1$. Here, $\Omega_r$, $\Omega_m$, and $\Omega_\Lambda$ denote the present-day radiation, matter, and cosmological-constant density parameters, respectively. Since the compression is calibrated within $\Lambda$CDM, this model provides the baseline case in which the compressed likelihood is expected to closely reproduce the constraints obtained from the full CMB analysis.

\subsection{$w$CDM}

The simplest extension of $\Lambda$CDM considered in this work replaces the cosmological constant with a dark-energy component characterized by a constant EoS parameter, $w_{\rm DE}(a)=w_0$ for all $a$.\footnote{Various works over the past years have examined constraints on $w_0$ in light of cosmological data. One of the most recent analyses finds $w_0=-1.013^{+0.038}_{-0.043}$ from a combination of SDSS BAO, \textit{PantheonPlus} unanchored SN, cosmic chronometers, and \textit{Planck} PR3 CMB likelihoods, fully consistent with the cosmological-constant scenario~\cite{Escamilla:2023oce}. A later analysis replacing SDSS BAO with DESI measurements finds $w_0=-1.049\pm0.024$, approximately $2\sigma$ below the cosmological-constant value~\cite{Sammut:2025eik}.} Its energy density evolves as $\rho_{\rm DE}(a)=\rho_{{\rm DE},0}a^{-3(1+w_0)}$, so that, for a spatially flat Universe,
\begin{equation}
E^2(a)=\Omega_r\, a^{-4}+\Omega_m\, a^{-3}+\Omega_{\rm DE}\, a^{-3(1+w_0)}.
\end{equation}
The $\Lambda$CDM limit is recovered for $w_0=-1$, while $w_0>-1$ and $w_0<-1$ correspond, respectively, to the quintessence-like and phantom regimes.
Importantly for our analysis, varying $w_0$ modifies the late-time expansion history while leaving the standard evolution of the matter densities unchanged. The physical matter abundance constrained around recombination therefore retains its usual relation to the present-day matter density, while changes in $H(a)$ modify the distance to last scattering and the subsequent growth of structure. At the perturbation level, dark-energy fluctuations and the modified late-time gravitational potentials affect the large-scale CMB anisotropies through the integrated Sachs-Wolfe (ISW) effect and alter the lensing of the primary CMB spectra. The primary acoustic physics at recombination, however, remains essentially unchanged.

\subsection{CPL dynamical dark energy}

We next consider a time-dependent dark-energy EoS described by the CPL parametrization~\cite{Chevallier:2000qy,Linder:2002et}\footnote{Several alternative, typically two-parameter, parametrizations of the EoS exist in the literature, see, e.g.,\ Refs.~\cite{Efstathiou:1999tm,Jassal:2004ej,Feng:2004ff,Gong:2005de,Barboza:2008rh,Ma:2011nc,Pantazis:2016nky,Escamilla-Rivera:2016qwv,Yang:2017alx,Pan:2017zoh,Yang:2018qmz,Capozziello:2020ctn,DiValentino:2020naf,Capozziello:2022jbw,Chaudhary:2023vxz,Singh:2023ryd,Adil:2023exv,Chaudhary:2023zzo,Dunsby:2023qpb,Rezaei:2024vtg,Kessler:2025kju,Shlivko:2025fgv,Cheng:2025lod,Lee:2025pzo,Fazzari:2025lzd}.} given by $w_{\rm DE}(a)=w_0+w_a(1-a)$. Here $w_0$ denotes the present-day value of the dark-energy EoS, while $w_a$ controls its evolution with scale factor.\footnote{Although the CPL form can be formally interpreted as a first-order expansion of $w_{\rm DE}(a)$ around the present epoch, it has also been shown to provide an accurate representation of the observables predicted by a broad class of physical quintessence models~\cite{Linder:2002et,Linder:2007wa,Corasaniti:2002vg,Linder:2005ne}. It can therefore be viewed more generally as a physics-motivated description of the dark-energy phase space spanned by $w_{\rm DE}(a)$ and $dw_{\rm DE}(a)/d\ln a$~\cite{Linder:2024rdj}; see, however, Ref.~\cite{Nesseris:2025lke}. For a recent critical assessment of cosmological constraints on CPL dark energy, see Ref.~\cite{Giare:2025pzu}.} The corresponding dark-energy density evolves as
\begin{equation}
\rho_{\rm DE}(a)
=
\rho_{{\rm DE},0}\,
a^{-3(1+w_0+w_a)}
\exp\left[-3w_a(1-a)\right],
\end{equation}
so that the background expansion is given by
\begin{equation}
\begin{aligned}
E^2(a)
={}&
\Omega_r\, a^{-4}
+
\Omega_m\, a^{-3}
\\
&+
\Omega_{\rm DE}\,
a^{-3(1+w_0+w_a)}
\exp\left[-3w_a(1-a)\right].
\end{aligned}
\end{equation}
The $\Lambda$CDM limit is recovered for $(w_0,w_a)=(-1,0)$, while $w_a=0$ reduces the model to $w$CDM. Recent DESI DR2 results show a preference for the region $w_0>-1$ and $w_a<0$ when BAO measurements are combined with CMB and SN data~\cite{DESI:2025zgx}.\footnote{This finding is supported across several dataset combinations and has been shown to be relatively robust to different analysis choices~\cite{Giare:2024oil}, although its interpretation and significance remain under debate~\cite{Sakr:2025daj,Colgain:2025fct,Keeley:2025rlg,Roy:2025cxk,Toomey:2025xyo}.}

As in $w$CDM, this extension modifies only the dark-energy sector while the standard scaling of the matter densities is preserved and the physical matter abundance constrained around recombination retains its usual relation to its present-day value. At the perturbation level, the evolving dark-energy component also affects the late-time gravitational potentials and the growth of structure, leaving signatures through the ISW effect and the lensing of the primary CMB spectra. These effects remain relatively limited for the smooth dark-energy component considered here because $c_{s,\rm DE}^2=1$ suppresses dark-energy clustering on sub-horizon scales. The acoustic physics responsible for the primary CMB anisotropies at recombination, however, remains essentially unchanged.\footnote{When the dark-energy EoS crosses the phantom divide, dark-energy fluctuations are evolved using the parametrized post-Friedmann (PPF) prescription~\cite{Fang:2008sn,Hu:2008zd}, which ensures a regular treatment through $w_{\rm DE}=-1$.}

\subsection{Sign-switching cosmological constant}

We next consider the so-called $\Lambda_{\rm s}$CDM model, in which the positive cosmological constant of $\Lambda$CDM is replaced by a vacuum-energy component that changes sign at a transition redshift $z=z_\dagger$~\cite{Akarsu:2021fol,Akarsu:2023mfb}. For $z>z_\dagger$, the vacuum-energy contribution is negative and corresponds to an anti-de Sitter vacuum, while for $z<z_\dagger$ it becomes positive with the same magnitude, corresponding to a de Sitter vacuum. Defining the corresponding transition scale factor as $a_\dagger \equiv1/(1+z_\dagger)$, the background expansion reads
\begin{equation}
E^2(a)
=
\Omega_r\, a^{-4}
+
\Omega_m\, a^{-3}
+
\Omega_{\Lambda_{\rm s}}\,
\mathrm{sgn}(a-a_\dagger),
\label{eq:Hlscdm}
\end{equation}
where $\mathrm{sgn}(x)$ denotes the signum function. The $\Lambda$CDM expansion history is recovered in the limit $z_\dagger\rightarrow\infty$.

The model was originally motivated by the graduated dark-energy scenario~\cite{Akarsu:2019hmw}, in which the dark-energy density evolves continuously and can change sign during the late-time expansion. Subsequent analyses showed that cosmological observations favor a relatively rapid transition~\cite{Akarsu:2022typ}, motivating the idealized $\Lambda_{\rm s}$CDM limit in which the sign change is taken to be abrupt.\footnote{Different microphysical realizations of such a phenomenological parametrization have been proposed, ranging from Casimir forces and thin-wall vacuum decay in string-inspired extra-dimensional scenarios~\cite{Anchordoqui:2023woo,Anchordoqui:2024gfa,Anchordoqui:2024dqc}, to type-II minimally modified gravity~\cite{Akarsu:2024qsi,Akarsu:2024eoo} and teleparallel $f(T)$ gravity~\cite{Akarsu:2024nas,Souza:2024qwd}. Depending on the underlying microphysics, additional degrees of freedom or modified perturbation dynamics may arise. Such effects are model dependent and are not considered in this work.} The $\Lambda_{\rm s}$CDM model has since been extensively studied~\cite{Paraskevas:2023itu,Anchordoqui:2023woo,Paraskevas:2024ytz,Akarsu:2024qsi,Anchordoqui:2024gfa,Akarsu:2024eoo,Yadav:2024duq,Anchordoqui:2024dqc,Akarsu:2024nas,Souza:2024qwd,Soriano:2025gxd,Akarsu:2025ijk,Bouhmadi-Lopez:2025ggl,Bouhmadi-Lopez:2025spo,Yadav:2025vpx,Akarsu:2025nns}, partly because of its ability to mildly alleviate the Hubble tension and the $S_8$ discrepancy for sufficiently low transition redshifts, $z_\dagger\lesssim2$~\cite{Akarsu:2021fol,Akarsu:2022typ,Akarsu:2023mfb}. However, like other late-time modifications of the expansion history, $\Lambda_{\rm s}$CDM remains subject to the general limitations of late-time solutions~\cite{Knox:2019rjx,Krishnan:2021dyb,Cai:2021weh,Cai:2022dkh,Raveri:2023zmr,Poulin:2024ken,Teixeira:2025czm,Pedrotti:2025ccw,Bansal:2026axl,Pedrotti:2026dwj,Sabogal:2026ipu}.

For the purposes of this work, we use the $\Lambda_{\rm s}$CDM model as a complementary test of CMB compression. The negative vacuum-energy contribution at $z>z_\dagger$ reduces the expansion rate relative to $\Lambda$CDM during part of the post-recombination evolution. Preserving the CMB angular acoustic scale then requires a compensating change at lower redshift, which can lead to the larger inferred values of $H_0$ discussed above. As in the $w$CDM and CPL cases, the physical matter density continues to scale as $\rho_m\propto a^{-3}$. Therefore, the physical matter abundance constrained around recombination retains its standard relation to its present-day value, while the primary acoustic physics at recombination remains unchanged. Unlike $w$CDM and CPL, however, $\Lambda_{\rm s}$CDM introduces a much sharper and more localized modification of the late-time expansion. At the perturbation level, on either side of the transition, the vacuum-energy component behaves as a cosmological constant and carries no independent perturbations.

\subsection{Interacting dark energy}

We next consider an interacting dark-energy (IDE) scenario, in which dark matter and dark energy are allowed to exchange energy and momentum through a non-gravitational interaction. Such models have been extensively investigated as simple extensions of the standard cosmological framework, both from a theoretical perspective and as possible phenomenological descriptions of cosmological tensions~\cite{Salvatelli:2013wra,Kumar:2016zpg, Caprini:2016qxs, Murgia:2016ccp, Zheng:2017asg, Kumar:2017dnp, DiValentino:2017iww,Kumar:2021eev,Gao:2021xnk,Pan:2023mie,Benisty:2024lmj,Yang:2020uga,Forconi:2023hsj,Pourtsidou:2016ico,DiValentino:2020vnx,DiValentino:2020leo,Nunes:2021zzi,Yang:2018uae,Zhang:2018mlj,vonMarttens:2019ixw,Lucca:2020zjb,Xiao:2021nmk,Wang:2024vmw,Gao:2022ahg,Zhai:2023yny,Joseph:2022khn,Bernui:2023byc,Becker:2020hzj,Hoerning:2023hks,Giare:2024ytc,Giare:2024smz,Mukhopadhyay:2020bml,Escamilla:2023shf,vanderWesthuizen:2023hcl,Silva:2024ift,DiValentino:2019ffd,Zhao:2022ycr,Li:2024qso,Pooya:2024wsq,Halder:2024uao,Castello:2023zjr,Yao:2023jau,Mishra:2023ueo,Nunes:2016dlj,Silva:2025hxw,Yang:2025uyv,vanderWesthuizen:2025rip,Zhang:2025dwu,Costa:2025kwt,Li:2025muv,Li:2025owk,Li:2026xaz,Zhai:2026uwr,Zhang:2026dtd}.

We focus on the widely studied interaction in which the energy-momentum transfer is proportional to the dark-energy density~\cite{Gavela:2009cy,Gavela:2010tm,Clemson:2011an}. At the covariant level, the conservation equations for the two dark components are modified according to~\cite{Gavela:2010tm}
\begin{equation}
\nabla_\nu T_{\rm dm}^{\mu\nu}
=
\frac{Q\, u_{\rm dm}^{\mu}}{a},
\qquad
\nabla_\nu T_{\rm DE}^{\mu\nu}
=
-\frac{Q\, u_{\rm dm}^{\mu}}{a},
\end{equation}
where $u_{\rm dm}^{\mu}$ is the dark-matter four-velocity and
\begin{equation}
Q=\xi \mathcal{H}\rho_{\rm DE}.
\label{eq:IDE_Q}
\end{equation}
Here $\mathcal{H}=aH$ is the conformal Hubble parameter and $\xi$ a dimensionless coupling constant. With this convention, $\xi<0$ corresponds to an energy-momentum transfer from dark matter to dark energy, while $\xi>0$ describes the opposite direction. Since the energy-momentum transfer is taken to be parallel to the dark-matter four-velocity, there is no momentum transfer in the dark-matter rest frame.

At the background level, the continuity equations become
\begin{align}
\rho_{\rm dm}' + 3\mathcal{H}\rho_{\rm dm}
&=
\xi\mathcal{H}\rho_{\rm DE},
\\
\rho_{\rm DE}'
+3\mathcal{H}(1+w_{\rm DE})\rho_{\rm DE}
&=
-\xi\mathcal{H}\rho_{\rm DE},
\end{align}
where a prime denotes a derivative with respect to conformal time. For constant $w_{\rm DE}(z)=w_0$, these equations can be integrated analytically to give
\begin{equation}
\rho_{\rm DE}(a)
=
\rho_{{\rm DE},0}\,
a^{-3(1+w_{0})-\xi},
\label{eq:IDE_rhode}
\end{equation}
and
\begin{equation}
\rho_{\rm dm}(a)
=
\rho_{{\rm dm},0}\,a^{-3}
+
\frac{\xi\,\rho_{{\rm DE},0}}
{3w_{0}+\xi}
\left[
a^{-3}
-
a^{-3(1+w_{0})-\xi}
\right].
\label{eq:IDE_rhodm}
\end{equation}
The dark-energy background evolution can therefore be expressed in terms of an effective EoS, $w_{\rm eff} = w_{0}+\xi/3 \simeq -1+\xi/3$, where in the last step we have used the vacuum-like EoS $w_{0}\simeq-1$ adopted in our analysis.\footnote{In the numerical implementation, we fix $w_{0}=-0.9999$ rather than exactly $-1$. This infinitesimal displacement from the cosmological-constant limit regularizes the early-time perturbation evolution while having a negligible effect on the background dynamics~\cite{Jackson:2009mz,Gavela:2009cy}.}

The interaction also modifies the evolution of linear perturbations. We evolve the perturbations in synchronous gauge and adopt a dark-energy rest-frame sound speed $c_{s,\rm DE}^2=1$. Defining $\delta_i\equiv\delta\rho_i/\rho_i$ and denoting by $\theta_i$ the velocity divergence of the $i$-th component, for the vacuum-like constant EoS adopted in our analysis the coupled dark-sector perturbation equations are
\begin{align}
\delta_{\rm DE}'=&
-\xi
\left(
\frac{k v_T}{3}
+\frac{h'}{6}
\right)
-6\mathcal{H}\delta_{\rm DE}
\nonumber\\
&-
6\xi\mathcal{H}^2
\frac{\theta_{\rm DE}}{k^2},
\label{eq:IDE_deltaDE}
\\[1ex]
\theta_{\rm DE}'
={}&
2\mathcal{H}\theta_{\rm DE}
+
\frac{k^2}{1+w_0}\delta_{\rm DE}
\nonumber\\
&+
\frac{\xi\mathcal{H}}{1+w_0}
\left(
2\theta_{\rm DE}-\theta_{\rm dm}
\right),
\label{eq:IDE_thetaDE}
\\[1ex]
\delta_{\rm dm}'
={}&
-\theta_{\rm dm}
-\frac{h'}{2}
+
\xi\mathcal{H}
\frac{\rho_{\rm DE}}{\rho_{\rm dm}}
\left(
\delta_{\rm DE}-\delta_{\rm dm}
\right)
\nonumber\\
&+
\xi
\frac{\rho_{\rm DE}}{\rho_{\rm dm}}
\left(
\frac{k v_T}{3}
+\frac{h'}{6}
\right),
\label{eq:IDE_deltadm}
\\[1ex]
\theta_{\rm dm}'
={}&
-\mathcal{H}\theta_{\rm dm}.
\label{eq:IDE_thetadm}
\end{align}
Here $h$ is the trace of the scalar metric perturbation, $k$ is the Fourier wavenumber, and $v_T$ denotes the center-of-mass velocity of the total fluid.

For the purposes of this work, we restrict our analysis to $\xi<0$, corresponding to an energy transfer from dark matter to dark energy. A key consequence of Eq.~\eqref{eq:IDE_rhodm} is that the dark-matter density no longer follows the standard scaling $\rho_{\rm dm}\propto a^{-3}$ throughout its evolution. Although the interaction becomes negligible at sufficiently high redshift and the local scaling asymptotically approaches $a^{-3}$, the present-day value $\rho_{{\rm dm},0}$ is no longer sufficient by itself to determine the dark-matter abundance at recombination through the usual $\Lambda$CDM relation. For the negative coupling considered here, extrapolating $\rho_{{\rm dm},0}$ backward assuming the standard $a^{-3}$ scaling underestimates the actual dark-matter abundance at recombination as a consequence of the cumulative transfer of energy from dark matter to dark energy at later times. This distinguishes IDE from the $w$CDM, CPL, and $\Lambda_{\rm s}$CDM scenarios considered above, in which the dark-energy sector modifies the expansion history while the standard evolution of matter is preserved. In addition, the full CMB spectra retain perturbation-level information associated with the coupled dark sector that is not explicitly represented in the compressed likelihood.

\subsection{Decaying dark matter into dark radiation}

We then consider a decaying dark-matter scenario in which a cold, pressureless dark-matter component (hereafter denoted by the subscript dcdm) decays into invisible relativistic particles that behave as dark-radiation (hereafter denoted by the subscript dr)~\cite{Audren:2014bca,Poulin:2016nat}.\footnote{For interesting studies of decaying dark matter into invisible relativistic particles and dark radiation, see, e.g., Refs.~\cite{Oguri:2003nn,Takahashi:2003iu,Ichiki:2004vi,DeLopeAmigo:2009dc,Peter:2010xe,Huo:2011nz,Wang:2012eka,Wang:2013rha,Audren:2014bca,Blackadder:2014wpa,Kalashev:2019xkw,Blackadder:2015uta,Chudaykin:2016yfk,Poulin:2016nat,Bringmann:2018jpr,Pandey:2019plg,Vattis:2019efj,Enqvist:2019tsa,Blinov:2020uvz,Haridasu:2020xaa,Clark:2020miy,FrancoAbellan:2020xnr,Nygaard:2020sow,FrancoAbellan:2021sxk,Hubert:2021khy,Lester:2021noa,Anchordoqui:2022gmw,Simon:2022ftd,Davari:2022uwd,Alvi:2022aam,Holm:2022eqq,Fuss:2022zyt,Holm:2022kkd,Tanimura:2023bkh,Nygaard:2023gel,Lester:2025hqt,Montandon:2025xpd,Yadav:2026dhg}.}

At the background level, the energy densities of the decaying dark matter and the produced dark radiation evolve according to
\begin{align}
\rho_{\rm dcdm}'
&=
-3\mathcal{H}\rho_{\rm dcdm}
-a\Gamma_{\rm dcdm}\rho_{\rm dcdm},
\\
\rho_{\rm dr}'
&=
-4\mathcal{H}\rho_{\rm dr}
+a\Gamma_{\rm dcdm}\rho_{\rm dcdm},
\end{align}
where $\Gamma_{\rm dcdm}$ is the constant decay rate defined with respect to proper time. The decaying dark-matter density therefore evolves as
\begin{equation}
\rho_{\rm dcdm}(a)
=
\rho_{{\rm dcdm},0}\,
a^{-3}
\exp\left[\Gamma_{\rm dcdm}\left(t_0-t(a)\right)\right],
\label{eq:DDM_rho_evolution}
\end{equation}
showing explicitly that its abundance at earlier times is larger than that obtained by extrapolating the present-day density with the standard $a^{-3}$ scaling. If the initial dark-radiation abundance is negligible, its subsequent evolution is entirely sourced by the decay of the dark-matter component,
\begin{equation}
\rho_{\rm dr}(a)
=
\frac{1}{a^4}
\int_{t_{\rm ini}}^{t(a)}
dt'\,
a^4(t')\,
\Gamma_{\rm dcdm}\,\rho_{\rm dcdm}(t').
\end{equation}
Thus, although each contribution produced by the decay subsequently redshifts as radiation, the total dark-radiation density does not follow a simple $a^{-4}$ scaling while it continues to be sourced. The $\Lambda$CDM limit is recovered for $\Gamma_{\rm dcdm}=0$.

The decay also modifies the perturbation evolution. We work in synchronous gauge comoving with the decaying dark-matter component. Since it remains pressureless, its perturbations obey
\begin{align}
\delta_{\rm dcdm}'
&=
-\frac{h'}{2},
\\
\theta_{\rm dcdm}'
&=
-\mathcal{H}\theta_{\rm dcdm},
\end{align}
so that the gauge condition $\theta_{\rm dcdm}=0$ is preserved throughout the evolution. In this gauge, the decay does not introduce an explicit source term in the evolution of the dark-matter density contrast. Its effects instead enter through the modified background abundance of the decaying dark matter, the perturbations of the daughter dark-radiation component, and the resulting modification of the metric perturbations. For the dark-radiation component, the first two moments of the corresponding Boltzmann hierarchy satisfy
\begin{align}
\delta_{\rm dr}'
={}&
-\frac{4}{3}
\left(
\theta_{\rm dr}
+\frac{h'}{2}
\right)
\nonumber\\
&+
a\Gamma_{\rm dcdm}
\frac{\rho_{\rm dcdm}}{\rho_{\rm dr}}
\left(
\delta_{\rm dcdm}-\delta_{\rm dr}
\right),
\\[1ex]
\theta_{\rm dr}'
={}&
\frac{k^2}{4}\delta_{\rm dr}
-k^2\sigma_{\rm dr}
\nonumber\\
&-
a\Gamma_{\rm dcdm}
\frac{3\rho_{\rm dcdm}}{4\rho_{\rm dr}}
\left(
\frac{4}{3}\theta_{\rm dr}
-\theta_{\rm dcdm}
\right),
\end{align}
where $\sigma_{\rm dr}$ denotes the dark-radiation anisotropic stress. The higher moments are evolved through the corresponding Boltzmann hierarchy.

Concretely, we use the standard implementation of this model in \texttt{CLASS}~\cite{Poulin:2016nat}. The sampled density parameter is
\begin{equation}
\omega_{\rm dcdmdr}
\equiv
\left(
\Omega_{{\rm dcdm}}
+
\Omega_{{\rm dr}}
\right)h^2,
\end{equation}
which specifies the combined present-day abundance of the decaying dark-matter component and the dark radiation produced by the decay. Since this quantity is defined at the present epoch whereas the background evolution is initialized at early times, \texttt{CLASS} determines the required initial abundance of the decaying dark-matter component through a shooting procedure.\footnote{For each pair $(\omega_{\rm dcdmdr},\Gamma_{\rm dcdm})$, the initial DCDM abundance is iteratively adjusted until the evolved solution reproduces the target value of $\omega_{\rm dcdmdr}$ at $z=0$, while the dark-radiation density is initialized according to its early-time asymptotic solution sourced by the decay.} Linear perturbations, and in particular the full collisionless Boltzmann hierarchy associated with dark radiation, are evolved consistently~\cite{Poulin:2016nat}.

For the purposes of this work, this model provides a second test of CMB compression in the presence of non-standard dark-matter properties. As in the IDE case, the background modifications become relevant primarily at late times. Unlike IDE, however, they are confined to the dark-matter sector.

\subsection{$\Lambda w$DM}

We now consider the $\Lambda w$DM model, in which dark energy remains a cosmological constant while dark matter is allowed to have a constant EoS, $w_{\rm dm}(z)\equiv p_{\rm dm}/\rho_{\rm dm}=w_{\rm dm,0}$.\footnote{For works discussing this possibility, see, e.g., Refs.~\cite{Muller:2004yb,Serra:2011jh,Kumar:2012gr,Avelino:2012jy,Cruz:2012ye,Wei:2013et,Xu:2013mqe,Yang:2013ria,Thomas:2016iav,Kopp:2016mhm,Kopp:2018zxp,Kumar:2019gfl,Liu:2024wne,Kumar:2025etf,Li:2025eqh,Li:2025dwz,Liu:2025mob,Abedin:2025dis,Wang:2025zri,Wang:2025rll,Yao:2025twv,Jiang:2026ngo}.} The standard $\Lambda$CDM model is recovered in the limit $w_{\rm dm,0}=0$.

At the background level, energy conservation gives
\begin{equation}
\rho_{\rm dm}(a)
=
\rho_{{\rm dm},0}\,
a^{-3(1+w_{\rm dm,0})}.
\label{eq:lwdm_background}
\end{equation}
Thus, for $w_{\rm dm,0}\neq0$, the dark-matter abundance at recombination is no longer related to its present-day value through the standard $\rho_{\rm dm}\propto a^{-3}$ scaling.

The evolution of dark-matter perturbations requires an additional prescription relating pressure and density fluctuations~\cite{Hu:1998kj}. We adopt a phenomenological fluid description with vanishing anisotropic stress and fix the dark-matter rest-frame sound speed to $c_{s,\rm dm}^2=0$.\footnote{For constant $w_{\rm dm}(z)=w_{\rm dm,0}$, the adiabatic sound speed is $c_{a,\rm dm}^2\equiv p_{\rm dm}'/\rho_{\rm dm}'=w_{\rm dm,0}$. Consequently, the choice $c_{s,\rm dm}^2=0$ describes an intrinsically non-adiabatic dark-matter fluid whenever $w_{\rm dm,0}\neq0$.} This eliminates pressure-gradient support and allows the dark-matter component to remain clustering despite its non-zero background EoS. In synchronous gauge, the corresponding Euler equation takes the standard form
\begin{equation}
\theta_{\rm dm}'=-\mathcal{H}\theta_{\rm dm}.
\end{equation}
The usual gauge choice comoving with dark matter, $\theta_{\rm dm}=0$, is therefore preserved. The density contrast instead obeys
\begin{equation}
\delta_{\rm dm}'
=
-\frac{1+w_{\rm dm,0}}{2}h'
+
3\mathcal{H}w_{\rm dm,0}\,\delta_{\rm dm}.
\label{eq:lwdm_delta}
\end{equation}
The perturbations therefore differ from those of standard CDM even though the rest-frame sound speed vanishes.

For the purposes of this work, $\Lambda w$DM provides a second direct test of CMB compression in a scenario where the dark-matter sector is modified while dark energy remains a cosmological constant. The main difference with the decaying-dark-matter scenario is that, while the effects of dark-matter decay become most relevant at late times, in $\Lambda w$DM the dark-matter evolution differs from the standard scaling throughout the entire cosmological history.

\subsection{$\Lambda w(a)$DM}
Finally, we generalize the constant-EoS model introduced above by allowing the dark-matter EoS to vary with time, while keeping dark energy as a cosmological constant. Specifically, we adopt
\begin{equation}
w_{\rm dm}(a)=w_{\rm dm,0}\,a,
\label{eq:lwadm_eos}
\end{equation}
where $w_{\rm dm,0}$ denotes the present-day EoS. This parametrization introduces no additional free parameters relative to the constant-EoS model, but describes a different evolution: $w_{\rm dm}(a)$ approaches zero at high redshift and becomes appreciable only at late times. The $\Lambda$CDM limit is again recovered for $w_{\rm dm,0}=0$.

Integrating the background continuity equation gives
\begin{equation}
\rho_{\rm dm}(a)
=
\rho_{{\rm dm},0}\,
a^{-3}
\exp\left[3w_{\rm dm,0}(1-a)\right].
\label{eq:lwadm_background}
\end{equation}
Unlike the constant-EoS case, the logarithmic slope of the dark-matter density therefore approaches the standard pressureless value at high redshift. However, its normalization encodes the cumulative effect of the nonzero EoS at later times with
\begin{equation}
\lim_{a\rightarrow0}
\frac{\rho_{\rm dm}(a)}
{\rho_{{\rm dm},0}a^{-3}}
=
e^{3w_{\rm dm,0}}.
\label{eq:lwadm_early_normalization}
\end{equation}
Thus, recovering the standard early-time scaling does not restore the usual relation between the present-day dark-matter density and its abundance around recombination.

The perturbations are treated with the same fluid prescription as in the $\Lambda w$DM model, with vanishing anisotropic stress and rest-frame sound speed $c_{s,\rm dm}^2=0$. The Euler equation remains unchanged, so that the synchronous-gauge condition comoving with dark matter, $\theta_{\rm dm}=0$, is preserved. In this gauge, the density contrast obeys
\begin{equation}
\delta_{\rm dm}'
=
-\frac{1+w_{\rm dm}(a)}{2}h'
+
3\mathcal{H}w_{\rm dm}(a)\,\delta_{\rm dm},
\label{eq:lwadm_delta}
\end{equation}
which generalizes Eq.~\eqref{eq:lwdm_delta} to the time-dependent EoS.\footnote{For a time-dependent EoS, the adiabatic sound speed is $c_{a,\rm dm}^2=w_{\rm dm}-w_{\rm dm}'/[3\mathcal{H}(1+w_{\rm dm})]$. The terms involving $w_{\rm dm}'$ in the general density-perturbation equation are proportional to $\theta_{\rm dm}$ and therefore vanish in the comoving gauge adopted here.}

For our purposes, this model provides a complementary test to the constant $w_{\rm dm}$ scenario. Comparing the two cases can highlight the differences between an early-time normalization offset from a persistent modification of the dark-matter redshift scaling.
%===================
\section{Results}
\label{sec:results}

We now present the cosmological constraints obtained using the datasets and methodology described in Sec.~\ref{sec:methodology} for the models introduced in Sec.~\ref{sec:models}. For each model, we first compare the marginalized constraints and two-dimensional posterior distributions obtained from dataset $A$ (i.e., full Planck-CMB+DESI+PantheonPlus) and dataset $B$ (i.e., compressed Planck-CMB+DESI+PantheonPlus). We then complement this comparison with the multidimensional posterior diagnostics introduced in Sec.~\ref{sec:methodology}.

Throughout this section, we focus exclusively on the quantitative agreement or disagreement between the two posterior distributions. Readers primarily interested in the overall pattern emerging from the model-by-model analysis may proceed directly to Sec.~\ref{sec:overall_pattern}. The physical origin of this pattern and its broader implications are instead discussed in Sec.~\ref{sec:discussion}.

The marginalized parameter constraints obtained from $A$ and $B$ for the different cosmological models are reported in Tables~\ref{tab:lcdm_constraints}--\ref{tab:LwaDM_constraints}, together with the corresponding one- and two-dimensional posterior distributions shown in Figures~\ref{fig:lcdm_results}--\ref{fig:LwaDM_results}. Unless otherwise stated, marginalized constraints are quoted at $68\%$ confidence level (CL), while one-tail bounds are given at $95\%$ CL. The multidimensional comparison of the posterior distributions obtained with \texttt{PostModes} is instead summarized in Table~\ref{tab:postmodes_summary}. Figure~\ref{fig:eigenvector_overlap} additionally shows the eigenvector-overlap matrices for $\Lambda$CDM, DCDMDR, $\Lambda w$DM, and $\Lambda w(a)$DM, which are discussed in the corresponding subsections below.

\subsection{$\Lambda$CDM}

\begin{figure}[ht!]
    \centering
    \includegraphics[width=0.9\linewidth]{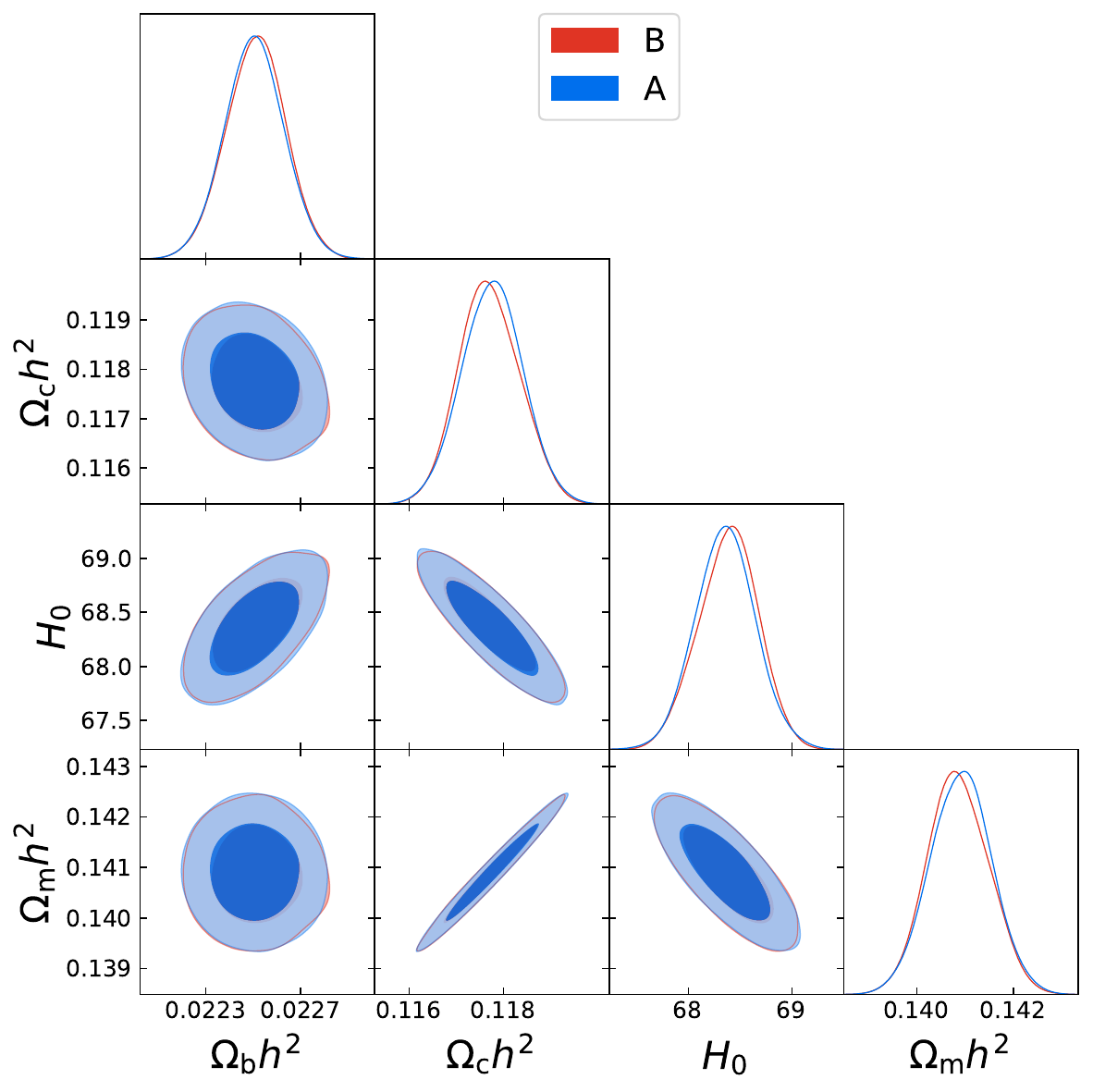}
    \caption{One- and two-dimensional posterior distributions obtained for the most informative parameters of the $\Lambda$CDM model. The datasets $A$ and $B$ differ by the inclusion of the full and compressed Planck CMB likelihoods, respectively.}
    \label{fig:lcdm_results}
\end{figure}

\begin{table}[ht!]
\centering
\begin{tabular}{lcc}
\hline
Parameter & $A$ & $B$ \\
\hline
$\Omega_{\mathrm b}h^2$ & $0.02251\pm0.00013$ & $0.02251\pm0.00013$ \\
$\Omega_{\mathrm c}h^2$ & $0.11776\pm0.00065$ & $0.11770^{+0.00061}_{-0.00069}$ \\
$H_0$ & $68.36\pm0.29$ & $68.39^{+0.30}_{-0.27}$ \\
$\Omega_{\mathrm m}$ & $0.3016\pm0.0037$ & $0.3012^{+0.0034}_{-0.0039}$ \\
$100\theta_{\mathrm s}$ & $1.04208\pm0.00027$ & $1.04208\pm0.00027$ \\
\hline
\end{tabular}
\caption{Marginalized constraints on the most informative parameters of the $\Lambda$CDM model. The datasets $A$ and $B$ differ by the inclusion of the full and compressed Planck CMB likelihoods, respectively.}
\label{tab:lcdm_constraints}
\end{table}

We begin with $\Lambda$CDM, which provides the reference test of our compression scheme. Since the compressed CMB likelihood is constructed from the full Planck posterior obtained within $\Lambda$CDM, the two analyses are expected to give nearly identical constraints. Figure~\ref{fig:lcdm_results} and Table~\ref{tab:lcdm_constraints} confirm this expectation after combining both CMB likelihoods with BAO and SN. The constraints on the physical baryon and dark-matter densities, $H_0$, $\Omega_{\rm m}$, and the angular acoustic scale are basically unchanged between $A$ and $B$, and the corresponding one- and two-dimensional posterior distributions are nearly indistinguishable.

The multidimensional comparison leads to the same conclusion. As summarized in Table~\ref{tab:postmodes_summary}, we find $S_{\rm CC}=0.070$, showing a negligible displacement of the posterior centers, while $\sqrt{\alpha}=0.994$ and $A_{\rm aniso}=0.038$ indicate essentially no change in the overall posterior scale or anisotropic deformation. The directional width ratios are similarly close to unity, spanning only $\sqrt{\rho_i}=0.970$--$1.016$. Finally, the $\Lambda$CDM overlap matrix shown in Figure~\ref{fig:eigenvector_overlap} (top-left panel) is essentially diagonal, confirming that the principal directions of the posterior are preserved.

\subsection{$w$CDM}

\begin{figure}[htpb!]
    \centering
    \includegraphics[width=0.9\linewidth]{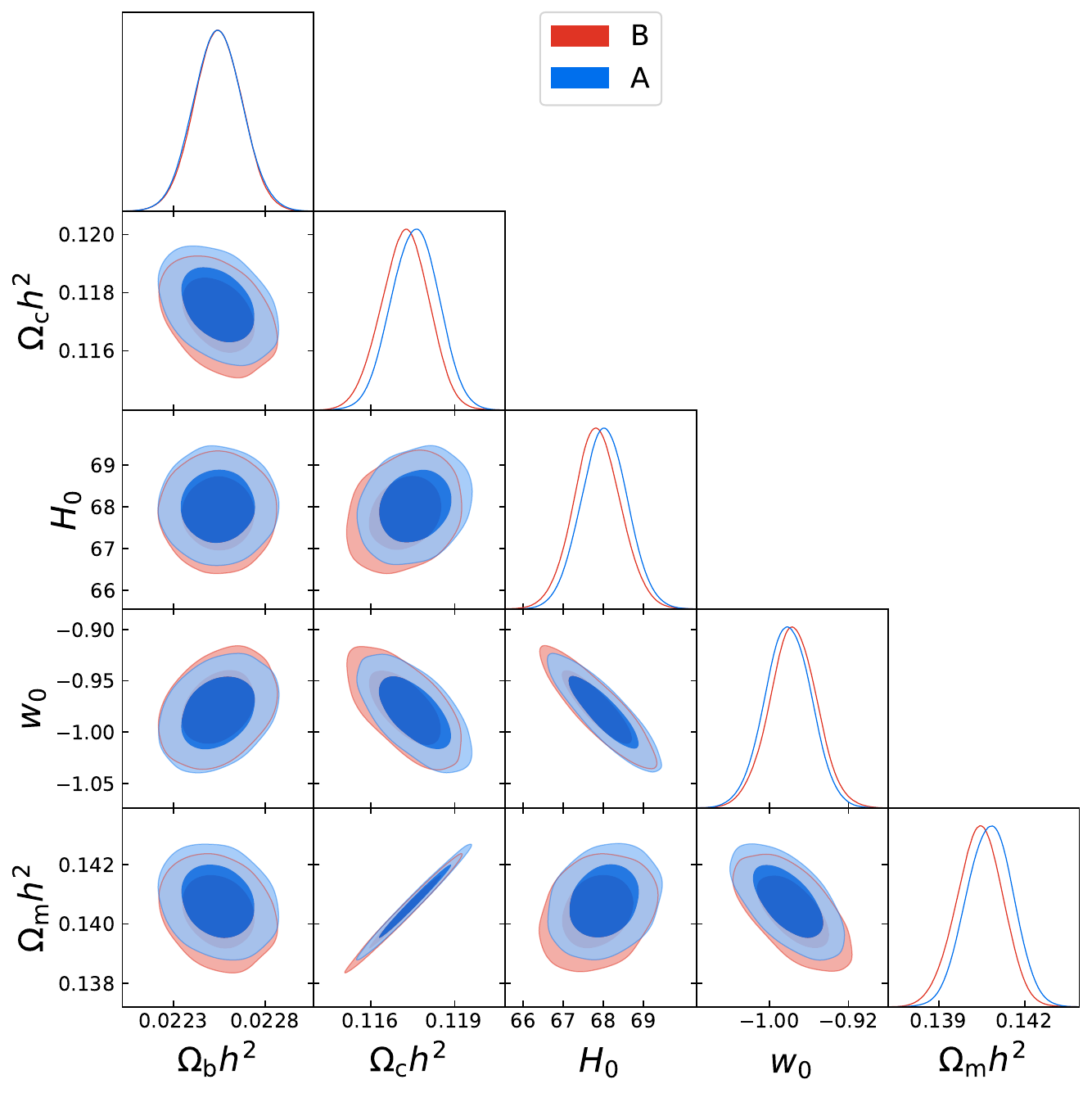}
    \caption{One- and two-dimensional posterior distributions obtained for the most informative  parameters of the $w$CDM model. The datasets $A$ and $B$ differ by the inclusion of the full and compressed Planck CMB likelihoods, respectively.}
    \label{fig:wcdm_results}
\end{figure}

\begin{table}[htpb!]
\centering
\begin{tabular}{lcc}
\hline
Parameter & $A$ & $B$ \\
\hline
$\Omega_{\mathrm b}h^2$ & $0.02254\pm0.00013$ & $0.02254\pm0.00013$ \\
$\Omega_{\mathrm c}h^2$ & $0.11758\pm0.00085$ & $0.11723\pm0.00085$ \\
$H_0$ & $68.02\pm0.58$ & $67.84\pm0.59$ \\
$w_{\rm DE}$ & $-0.981\pm0.024$ & $-0.976\pm0.024$ \\
$\Omega_{\mathrm m}$ & $0.3043\pm0.0051$ & $0.3051\pm0.0052$ \\
$100\theta_{\mathrm s}$ & $1.04208\pm0.00027$ & $1.04212\pm0.00028$ \\
\hline
\end{tabular}
\caption{Marginalized constraints on the most informative parameters of the $w$CDM model. The datasets $A$ and $B$ differ by the inclusion of the full and compressed Planck CMB likelihoods, respectively.}
\label{tab:wcdm_constraints}
\end{table}

The agreement remains excellent when a constant dark-energy EoS is introduced. As shown in Figure~\ref{fig:wcdm_results} and Table~\ref{tab:wcdm_constraints}, the marginalized constraints obtained from $A$ and $B$ are very similar. In particular, the full-CMB analysis gives $w_0=-0.981\pm0.024$, compared with $w_0=-0.976\pm0.024$ from the compressed-CMB analysis, with essentially unchanged uncertainty. The constraints on the remaining cosmological parameters are likewise closely reproduced, and the corresponding one- and two-dimensional posterior distributions largely overlap.

The multidimensional comparison further confirms the close agreement between the two analyses. The posterior centers differ only slightly, with $S_{\rm CC}=0.338$, while $\sqrt{\alpha}=1.004$ indicates essentially no overall rescaling and $A_{\rm aniso}=0.057$ remains small. The directional standard-deviation ratios are correspondingly close to unity, spanning $\sqrt{\rho_i}=0.962$--$1.039$. We also find no significant mixing between the principal directions, with the eigenvector-overlap matrix remaining nearly diagonal. Overall, the scale and directional structure of the two posteriors are very well preserved.

\subsection{CPL dynamical dark energy}

\begin{figure}[htpb!]
    \centering
    \includegraphics[width=0.99\linewidth]{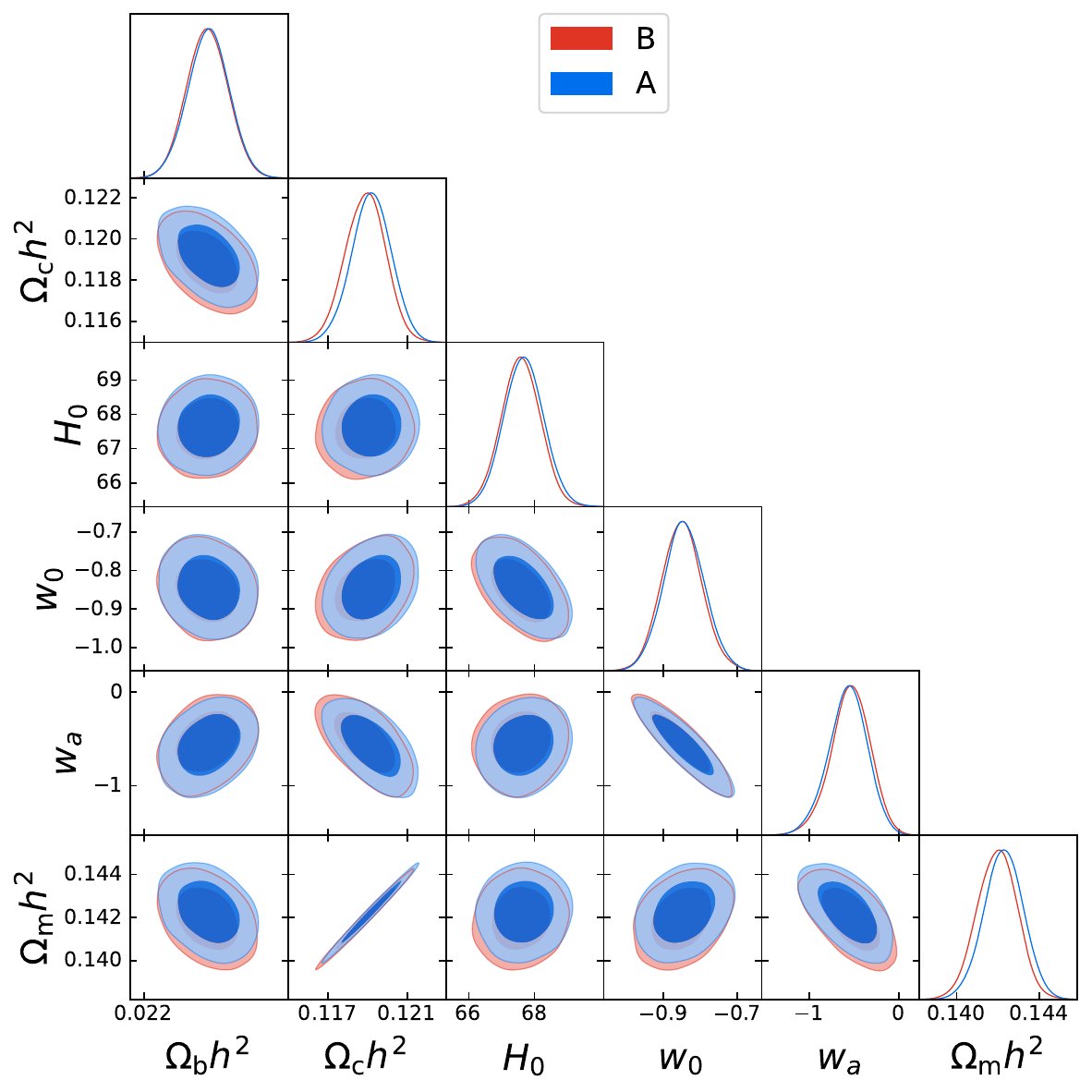}
    \caption{One- and two-dimensional posterior distributions obtained for the most informative  parameters of the CPL model. The datasets $A$ and $B$ differ by the inclusion of the full and compressed Planck CMB likelihoods, respectively.}
    \label{fig:CPL_results}
\end{figure}

A similarly close agreement is obtained for the two-parameter CPL model. As shown in Figure~\ref{fig:CPL_results} and Table~\ref{tab:CPL_constraints}, the two analyses select essentially the same region of the $(w_0,w_a)$ plane. The full-CMB likelihood gives $w_0=-0.844\pm0.055$ and $w_a=-0.57^{+0.23}_{-0.20}$, while the compressed likelihood gives $w_0=-0.850\pm0.055$ and $w_a=-0.54^{+0.23}_{-0.20}$. The constraints on the remaining cosmological parameters are likewise closely reproduced, and the corresponding one- and two-dimensional posterior distributions show a high degree of overlap.

The multidimensional comparison further confirms this agreement. The posterior centers differ only slightly, with $S_{\rm CC}=0.306$, while $\sqrt{\alpha}=0.998$ indicates essentially no overall rescaling and $A_{\rm aniso}=0.053$ remains small. The directional width ratios are also very close to unity, spanning $\sqrt{\rho_i}=0.966$--$1.035$. As in the $w$CDM case, the overall scale and directional structure of the posterior are very well preserved.

\begin{table}[htpb!]
\centering
\begin{tabular}{lcc}
\hline
Parameter & $A$ & $B$ \\
\hline
$\Omega_{\mathrm b}h^2$ & $0.02243\pm0.00014$ & $0.02243\pm0.00014$ \\
$\Omega_{\mathrm c}h^2$ & $0.1192\pm0.0010$ & $0.1189\pm0.0010$ \\
$H_0$ & $67.67\pm0.60$ & $67.58\pm0.59$ \\
$w_0$ & $-0.844\pm0.055$ & $-0.850\pm0.055$ \\
$w_a$ & $-0.57^{+0.23}_{-0.20}$ & $-0.54^{+0.23}_{-0.20}$ \\
$\Omega_{\mathrm m}$ & $0.3107\pm0.0058$ & $0.3109\pm0.0056$ \\
$100\theta_{\mathrm s}$ & $1.04192\pm0.00029$ & $1.04196\pm0.00029$ \\
\hline
\end{tabular}
\caption{Marginalized constraints on the most informative parameters of the CPL model. The datasets $A$ and $B$ differ by the inclusion of the full and compressed Planck CMB likelihoods, respectively.}
\label{tab:CPL_constraints}
\end{table}

\subsection{Sign-switching cosmological constant}

\begin{figure}[ht!]
    \centering
    \includegraphics[width=0.99\linewidth]{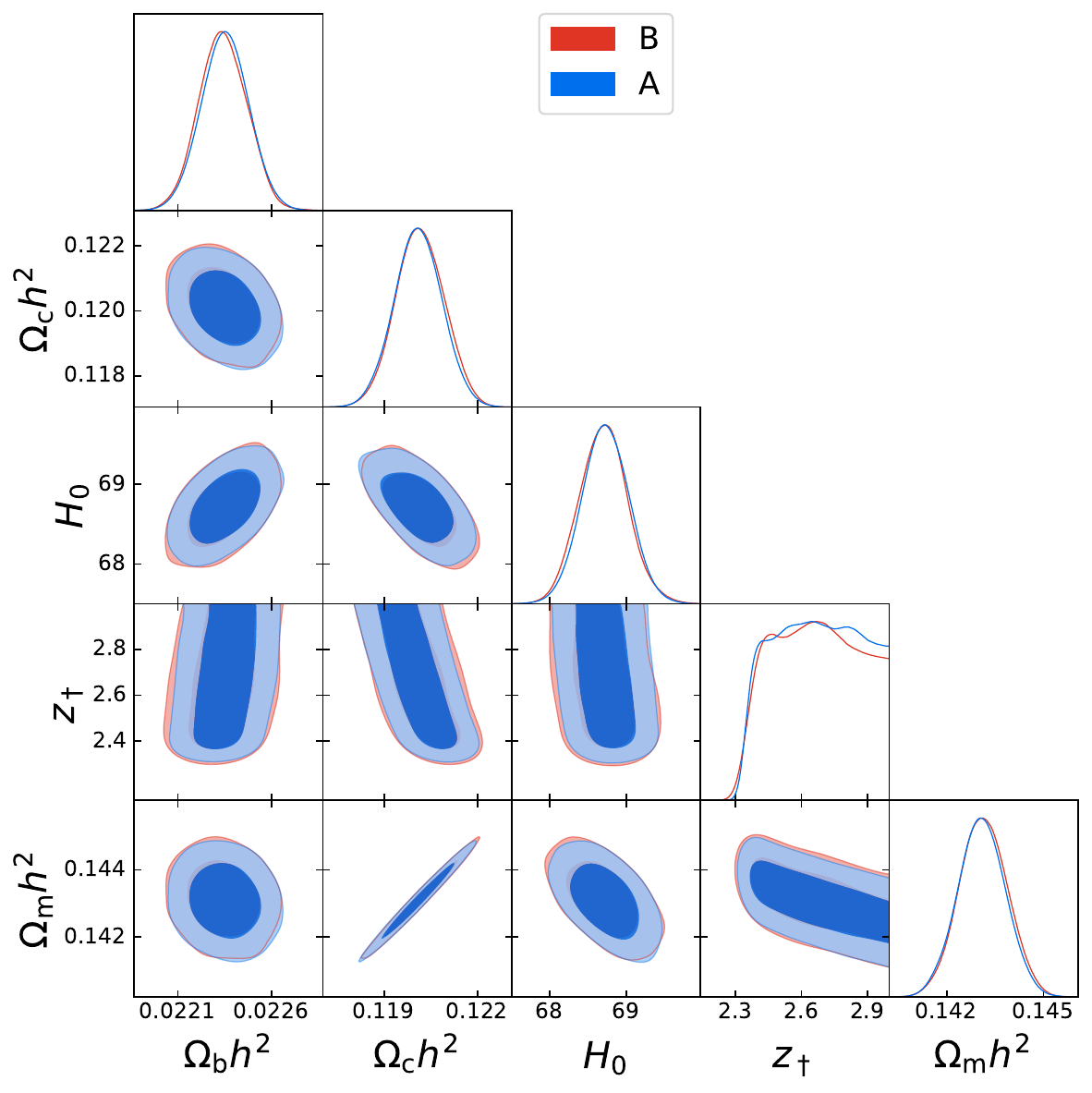}
    \caption{One- and two-dimensional posterior distributions obtained for the most informative  parameters of the $\Lambda_{\rm s}$CDM model. The datasets $A$ and $B$ differ by the inclusion of the full and compressed Planck CMB likelihoods, respectively.}
    \label{fig:LsCDM_results}
\end{figure}

\begin{table}[ht!]
\centering
\begin{tabular}{lcc}
\hline
Parameter & $A$ & $B$ \\
\hline
$\Omega_{\mathrm b}h^2$ & $0.02235\pm0.00013$ & $0.02234\pm0.00013$ \\
$\Omega_{\mathrm c}h^2$ & $0.12009\pm0.00076$ & $0.12015\pm0.00077$ \\
$H_0$ & $68.73\pm0.31$ & $68.70\pm0.32$ \\
$z_\dagger$ & $2.67\pm0.19$ & $2.66^{+0.17}_{-0.25}$ \\
$\Omega_{\mathrm m}$ & $0.3029\pm0.0038$ & $0.3033\pm0.0039$ \\
$100\theta_{\mathrm s}$ & $1.04184\pm0.00028$ & $1.04184\pm0.00028$ \\
\hline
\end{tabular}
\caption{Marginalized constraints on the most informative parameters of the $\Lambda_{\rm s}$CDM model. The datasets $A$ and $B$ differ by the inclusion of the full and compressed Planck CMB likelihoods, respectively.}
\label{tab:LsCDM_constraints}
\end{table}

The agreement remains excellent for the $\Lambda_{\rm s}$CDM model. As shown in Figure~\ref{fig:LsCDM_results} and Table~\ref{tab:LsCDM_constraints}, the transition redshift is constrained to $z_\dagger=2.67\pm0.19$ with the full CMB and $z_\dagger=2.66^{+0.17}_{-0.25}$ with the compressed likelihood. The constraints on the remaining cosmological parameters are likewise closely reproduced, and the corresponding one- and two-dimensional posterior distributions show almost complete overlap.

The multidimensional comparison confirms that the posterior centers are nearly coincident, with $S_{\rm CC}=0.074$, while $\sqrt{\alpha}=1.005$ indicates essentially no overall rescaling. $A_{\rm aniso}=0.029$ is the smallest among the extended models considered. The directional width ratios remain very close to unity, spanning $\sqrt{\rho_i}=0.994$--$1.030$. We also find no mixing between the principal directions, with the eigenvector-overlap matrix remaining basically diagonal.

\subsection{Interacting dark energy}

\begin{figure}[ht!]
    \centering
    \includegraphics[width=0.99\linewidth]{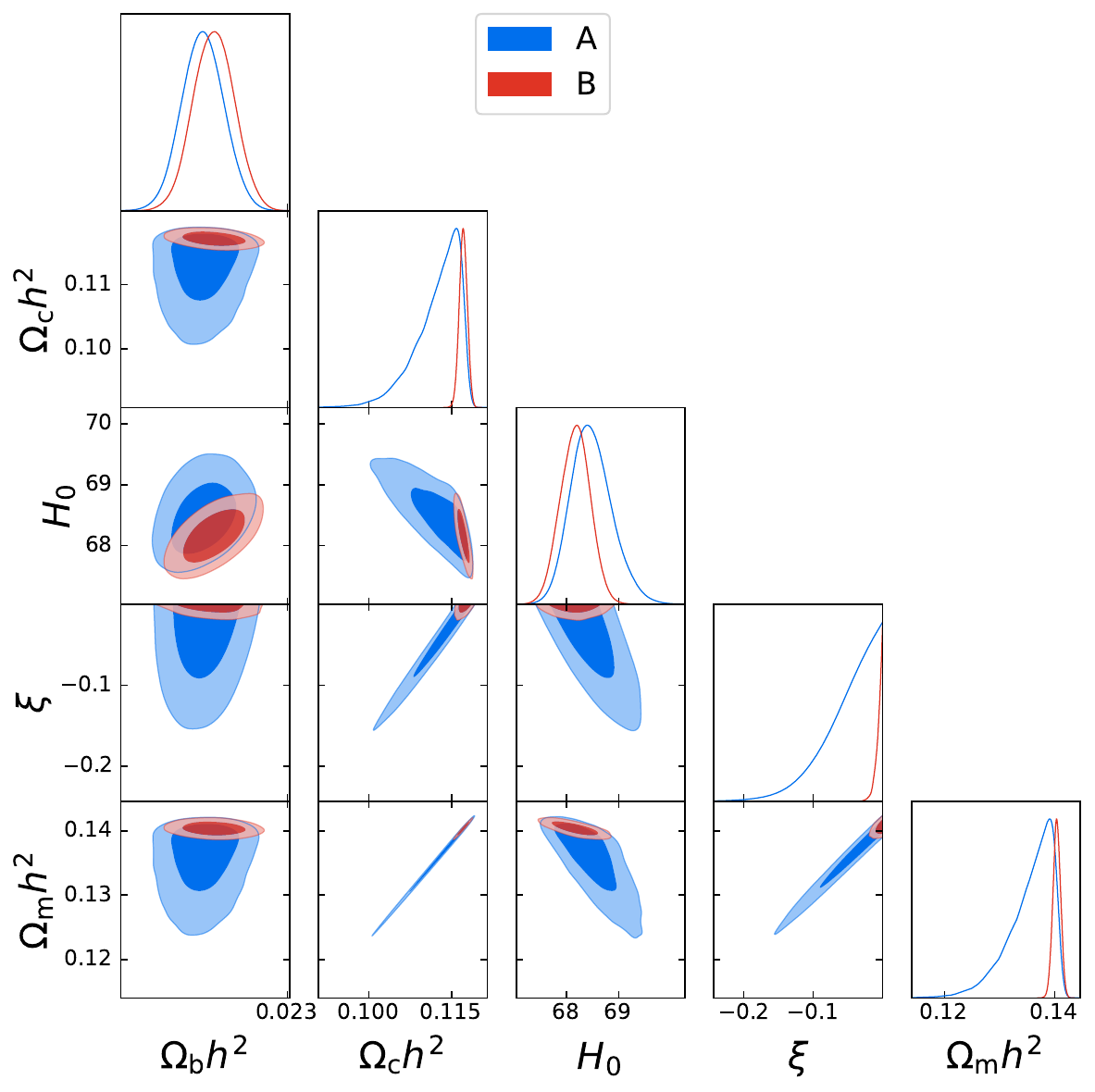}
    \caption{One- and two-dimensional posterior distributions obtained for the most informative  parameters of the IDE model. The datasets $A$ and $B$ differ by the inclusion of the full and compressed Planck CMB likelihoods, respectively.}
    \label{fig:IDE_results}
\end{figure}

\begin{table}[ht!]
\centering
\begin{tabular}{lcc}
\hline
Parameter & $A$ & $B$ \\
\hline
$\Omega_{\mathrm b}h^2$ & $0.02249\pm0.00013$ & $0.02256\pm0.00012$ \\
$\Omega_{\mathrm c}h^2$ & $0.1125^{+0.0050}_{-0.0018}$ & $0.11714\pm0.00072$ \\
$H_0$ & $68.48^{+0.34}_{-0.43}$ & $68.17\pm0.29$ \\
$\xi$ & $>-0.124$ & $>-0.0150$ \\
$\Omega_{\mathrm m}$ & $0.289^{+0.014}_{-0.0065}$ & $0.3021\pm0.0038$ \\
$100\theta_{\mathrm s}$ & $1.04208\pm0.00027$ & $1.04212\pm0.00028$ \\
\hline
\end{tabular}
\caption{Marginalized constraints on the most informative parameters of the IDE model. The datasets $A$ and $B$ differ by the inclusion of the full and compressed Planck CMB likelihoods, respectively.}
\label{tab:IDE_constraints}
\end{table}

As shown in Figure~\ref{fig:IDE_results} and Table~\ref{tab:IDE_constraints}, IDE provides the first clear case in which the compressed CMB likelihood fails to reproduce the cosmological constraints obtained from the full CMB likelihood, with substantial differences in both the marginalized constraints and their correlations. 

The most striking difference concerns the interaction parameter $\xi$. The full-CMB analysis gives $\xi>-0.124$, whereas the compressed likelihood yields the much tighter constraint $\xi>-0.0150$, reducing the allowed magnitude of the negative coupling by nearly an order of magnitude. The physical dark-matter density also changes from $\Omega_{\rm c}h^2=0.1125^{+0.0050}_{-0.0018}$ for $A$ to $0.11714\pm0.00072$ for $B$. A similarly substantial change is found in the matter density parameter, which shifts from $\Omega_{\rm m}=0.289^{+0.014}_{-0.0065}$ to $0.3021\pm0.0038$. Smaller differences are present in $H_0$ and $\Omega_{\rm b}h^2$. 

The two-dimensional posterior distributions confirm that the disagreement is concentrated primarily in projections involving $\xi$ and the matter-density parameters, where the extended degeneracy allowed by the full CMB is substantially reduced in the compressed analysis.

The multidimensional comparison shows that this difference is not simply a displacement between otherwise similar posteriors. We find $S_{\rm CC}=1.163$, indicating an appreciable separation of the posterior centers in the covariance-combined metric. At the same time, $\sqrt{\alpha}=0.573$ shows that the geometric-mean standard deviation of $B$ is only about $57\%$ of that of $A$. This contraction is highly direction dependent, as quantified by $A_{\rm aniso}=1.870$, which is dramatically larger than in the models considered above. The most strongly affected generalized mode has $\sqrt{\rho_{\min}}=0.114$, corresponding to a standard deviation approximately $8.8$ times smaller in $B$ than in $A$. The opposite extreme remains nearly unchanged, with $\sqrt{\rho_{\max}}=1.055$. The compressed likelihood therefore does not produce a uniform increase in constraining power, but instead strongly contracts particular combinations of parameters while leaving other directions almost unaffected. This strong anisotropic deformation ultimately leads to the substantially incorrect marginalized constraints discussed above.

\subsection{Decaying dark matter into dark radiation}

\begin{figure}[ht!]
    \centering
    \includegraphics[width=0.99\linewidth]{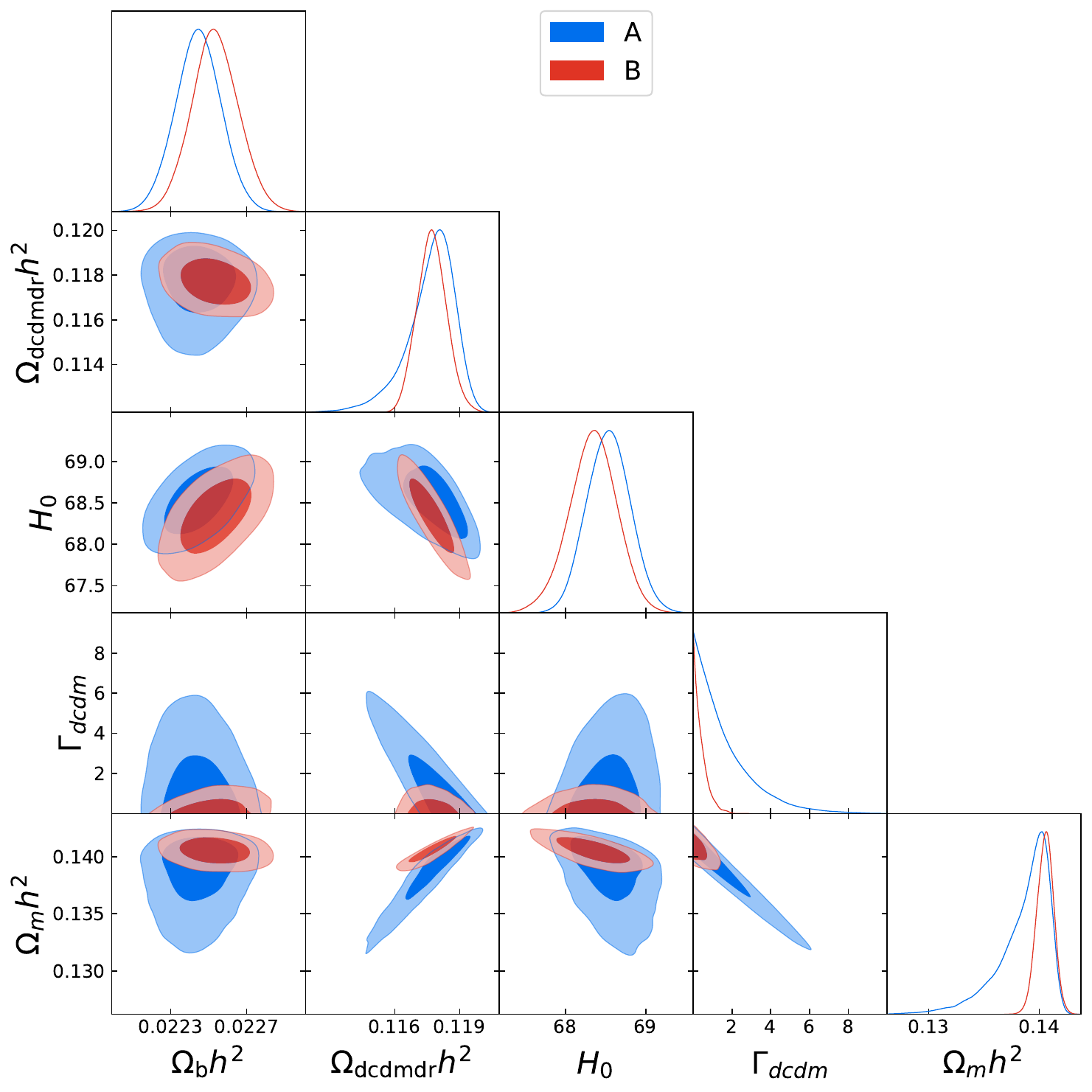}
    \caption{One- and two-dimensional posterior distributions obtained for the most informative  parameters of the DCDMDR model. The datasets $A$ and $B$ differ by the inclusion of the full and compressed Planck CMB likelihoods, respectively.}
    \label{fig:DCDMDR_results}
\end{figure}

\begin{table}[ht!]
\centering
\begin{tabular}{lcc}
\hline
Parameter & $A$ & $B$ \\
\hline
$\Omega_{\mathrm b}h^2$ & $0.02245\pm0.00012$ & $0.02254\pm0.00012$ \\
$\Omega_{\mathrm{dcdmdr}}h^2$ & $0.1177^{+0.0013}_{-0.00070}$ & $0.11771\pm0.00068$ \\
$H_0$ & $68.54\pm0.27$ & $68.34\pm0.30$ \\
$\Gamma_{\mathrm{dcdm}}$ & $<4.61$ & $<1.13$ \\
$\Omega_{\mathrm m}$ & $0.2952^{+0.0072}_{-0.0040}$ & $0.3009\pm0.0039$ \\
$100\theta_{\mathrm s}$ & $1.04199\pm0.00027$ & $1.04212\pm0.00029$ \\
\hline
\end{tabular}
\caption{Marginalized constraints on the most informative parameters of the DCDMDR model. The datasets $A$ and $B$ differ by the inclusion of the full and compressed Planck CMB likelihoods, respectively.}
\label{tab:DCDMDR_constraints}
\end{table}

The decaying dark matter scenario provides a second case in which the compressed and full-CMB analyses differ substantially. 

As shown in Figure~\ref{fig:DCDMDR_results} and Table~\ref{tab:DCDMDR_constraints}, the most pronounced difference concerns the decay rate. The full-CMB analysis gives $\Gamma_{\rm dcdm}<4.61$, whereas the compressed likelihood yields a substantially tighter constraint $\Gamma_{\rm dcdm}<1.13$. Despite the large difference in the allowed decay rate, the sampled combined density $\Omega_{\rm dcdmdr}h^2$ remains nearly unchanged. By contrast, the matter density parameter changes appreciably, from $\Omega_{\rm m}=0.2952^{+0.0072}_{-0.0040}$ for $A$ to $0.3009\pm0.0039$ for $B$, while the differences in $H_0$ and $\Omega_{\rm b}h^2$ remain comparatively small. The two-dimensional posterior distributions show that the compressed likelihood substantially restricts the region extending toward larger decay rates and modifies its correlations with the matter-density parameters.

The multidimensional comparison shows that this difference is accompanied by a substantial deformation of the posterior geometry. We find $S_{\rm CC}=0.977$, while $\sqrt{\alpha}=0.737$ indicates an overall contraction of the compressed posterior. As in IDE, this contraction is strongly direction dependent, with $A_{\rm aniso}=1.380$. The most strongly contracted generalized mode has $\sqrt{\rho_{\min}}=0.224$, corresponding to a standard deviation $\sim4.5$ times smaller in $B$ than in $A$. At the opposite extreme, $\sqrt{\rho_{\max}}=1.212$, so that one direction is instead about $21\%$ broader in the compressed posterior. As for IDE, the difference between $A$ and $B$ cannot be described by a uniform rescaling of the posterior widths.

The eigenvector-overlap matrix shown in the top-right panel of Figure~\ref{fig:eigenvector_overlap} reveals that the pronounced off-diagonal pattern is predominantly associated with a reordering of closely aligned modes rather than with a substantial rotation of the principal directions. Geometrically, the posterior therefore retains nearly the same principal axes, but their relative widths change sufficiently that their ordering by eigenvalue is altered between $A$ and $B$. After accounting for this reordering, the dominant mode correspondences have overlaps of $\sim0.96$--$1.00$, with only modest residual mixing. The dominant geometric effect is therefore a redistribution of the posterior widths among nearly preserved parameter combinations, rather than a major reorientation of the posterior eigensystem. This behavior is reflected in the marginalized constraints, where some parameters are more tightly constrained in $B$ than in $A$, while the overall correlation structure remains largely preserved.

\subsection{$\Lambda w$DM}

\begin{figure}[ht!]
    \centering
    \includegraphics[width=0.99\linewidth]{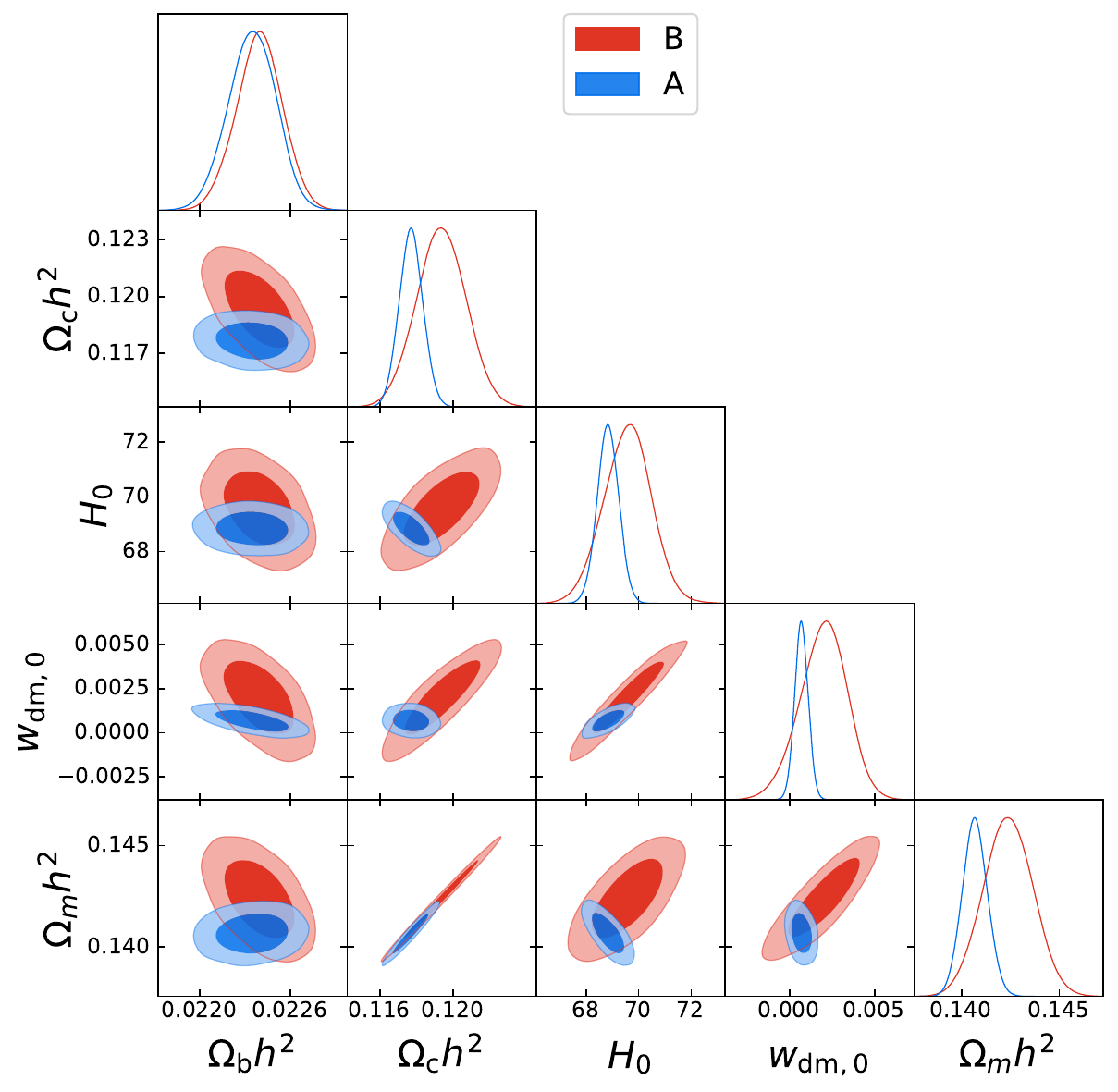}
    \caption{One- and two-dimensional posterior distributions obtained for the most informative parameters of the $\Lambda w$DM model. The datasets $A$ and $B$ differ by the inclusion of the full and compressed Planck CMB likelihoods, respectively.}
    \label{fig:LwDM_results}
\end{figure}

\begin{table}[ht!]
\centering
\begin{tabular}{lcc}
\hline
Parameter & $A$ & $B$ \\
\hline
$\Omega_{\mathrm b}h^2$ & $0.02234\pm0.00016$ & $0.02239\pm0.00015$ \\
$\Omega_{\mathrm c}h^2$ & $0.11769\pm0.00065$ & $0.1193\pm0.0013$ \\
$H_0$ & $68.83\pm0.41$ & $69.57\pm0.90$ \\
$w_{\mathrm{dm},0}$ & $0.00069\pm0.00040$ & $0.0020^{+0.0015}_{-0.0013}$ \\
$\Omega_{\mathrm m}$ & $0.2969\pm0.0046$ & $0.2943\pm0.0062$ \\
$100\theta_{\mathrm s}$ & $1.04193\pm0.00029$ & $1.04193\pm0.00030$ \\
\hline
\end{tabular}
\caption{Marginalized constraints on the most informative parameters of the $\Lambda w$DM model. The datasets $A$ and $B$ differ by the inclusion of the full and compressed Planck CMB likelihoods, respectively.}
\label{tab:LwDM_constraints}
\end{table}

A qualitatively different change in constraining power is found for the constant EoS $\Lambda w$DM model. As shown in Figure~\ref{fig:LwDM_results} and Table~\ref{tab:LwDM_constraints}, the full-CMB analysis gives $w_{\rm dm,0}=0.00069\pm0.00040$, whereas the compressed likelihood gives $w_{\rm dm,0}=0.0020^{+0.0015}_{-0.0013}$. The compressed analysis therefore shifts the preferred region toward more positive values and weakens the marginalized constraint by a factor of $\sim3.5$. Unlike the IDE and DCDMDR cases, the most pronounced difference is a loss of constraining power on the additional model parameter.

The physical dark matter density also changes, from $\Omega_{\rm c}h^2=0.11769\pm0.00065$ for $A$ to $0.1193\pm0.0013$ for $B$, while $H_0$ shifts from $68.83\pm0.41$ to $69.57\pm0.90$. Both parameters are less tightly constrained by the compressed likelihood. By contrast, the physical baryon density and the angular acoustic scale are closely reproduced. The two-dimensional posterior distributions also show that the compressed analysis produces stronger degeneracies between $w_{\rm dm,0}$, $H_0$, and the matter-density parameters as well as noticeable changes in their correlations.

The multidimensional comparison confirms that these differences cannot be described as a displacement between otherwise similar posteriors. As summarized in Table~\ref{tab:postmodes_summary}, we find $S_{\rm CC}=1.372$, while $\sqrt{\alpha}=1.404$ indicates an overall increase of approximately $40\%$ in the geometric-mean posterior width of $B$ relative to $A$. This broadening is strongly direction dependent, with $A_{\rm aniso}=1.476$. The broadest generalized mode has $\sqrt{\rho_{\max}}=4.924$, corresponding to a standard deviation almost five times larger in the compressed posterior. At the opposite extreme, we get $\sqrt{\rho_{\min}}=0.738$, corresponding to a standard deviation $\sim 26\%$ narrower in $B$ than in $A$. The overall broadening therefore masks a substantial redistribution of posterior widths across different parameter combinations.

The eigenvector-overlap matrix shown in the bottom-left panel of Figure~\ref{fig:eigenvector_overlap} reveals substantial mixing among the principal directions. In contrast to the predominantly reordered modes observed in DCDMDR, here a permutation of the mode indices does not recover a closely aligned basis. Even after accounting for the best correspondence between modes, the matched overlaps span $0.59$--$0.76$ in the common reference-standardized metric. The principal axes therefore undergo a substantial reorientation in addition to the changes in directional widths.

\subsection{$\Lambda w(a)$DM}

\begin{figure}[ht!]
    \centering
    \includegraphics[width=0.99\linewidth]{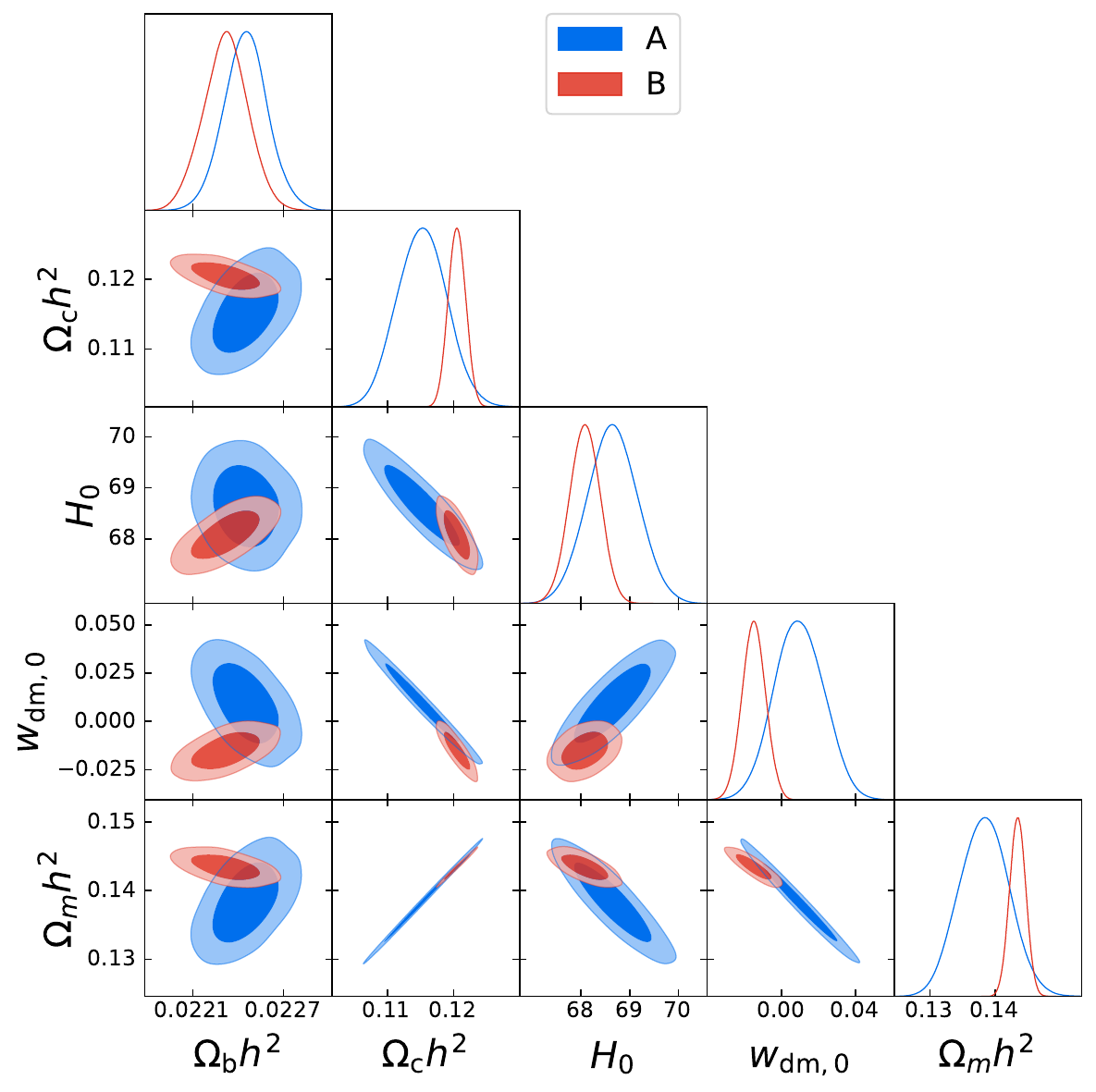}
    \caption{One- and two-dimensional posterior distributions obtained for the most informative parameters of the $\Lambda w(a)$DM model. The datasets $A$ and $B$ differ by the inclusion of the full and compressed Planck CMB likelihoods, respectively.}
    \label{fig:LwaDM_results}
\end{figure}

\begin{table}[ht!]
\centering
\begin{tabular}{lcc}
\hline
Parameter & $A$ & $B$ \\
\hline
$\Omega_{\mathrm b}h^2$ & $0.02245\pm0.00015$ & $0.02232\pm0.00015$ \\
$\Omega_{\mathrm c}h^2$ & $0.1152\pm0.0037$ & $0.1205\pm0.0013$ \\
$H_0$ & $68.65\pm0.52$ & $68.08\pm0.31$ \\
$w_{\mathrm{dm},0}$ & $0.009\pm0.013$ & $-0.0152\pm0.0063$ \\
$\Omega_{\mathrm m}$ & $0.294\pm0.012$ & $0.3096\pm0.0050$ \\
$100\theta_{\mathrm s}$ & $1.04202\pm0.00029$ & $1.04181\pm0.00030$ \\
\hline
\end{tabular}
\caption{Marginalized constraints on the most informative parameters of the $\Lambda w(a)$DM model. The datasets $A$ and $B$ differ by the inclusion of the full and compressed Planck CMB likelihoods, respectively.}
\label{tab:LwaDM_constraints}
\end{table}

The most prominent discrepancy between the two CMB prescriptions is found for $\Lambda w(a)$DM. 

As shown in Figure~\ref{fig:LwaDM_results} and Table~\ref{tab:LwaDM_constraints}, the compressed likelihood changes both the preferred parameter values and their uncertainties. The full-CMB analysis gives $w_{\rm dm,0}=0.009\pm0.013$, consistent with the cold-dark-matter limit, whereas the compressed likelihood gives $w_{\rm dm,0}=-0.0152\pm0.0063$, shifting the preferred region toward negative values while reducing the marginalized uncertainty by approximately a factor of two. This produces a spurious $\sim2.4\sigma$ preference for a non-zero dark-matter EoS, despite the full-CMB analysis being fully consistent with $w_{\rm dm,0}=0$.

We also observe a substantial shift in the physical dark-matter density, from $\Omega_{\rm c}h^2=0.1152\pm0.0037$ for $A$ to $0.1205\pm0.0013$ for $B$. Similarly, $\Omega_{\rm m}$ changes from $0.294\pm0.012$ to $0.3096\pm0.0050$. The corresponding two-dimensional posterior distributions show that $A$ and $B$ select noticeably different regions in projections involving $w_{\rm dm}$ and the matter-density parameters.

\begin{table*}[ht!]
\centering
\begin{tabular}{
l
@{\hspace{0.8cm}}c
@{\hspace{0.6cm}}c
@{\hspace{0.6cm}}c
@{\hspace{0.6cm}}c
@{\hspace{0.6cm}}c
@{\hspace{0.6cm}}c
}
\hline
\textbf{Model} & \boldmath$N$ & \boldmath$S_{\rm CC}$ & \boldmath$\sqrt{\alpha}$ &
\boldmath$A_{\rm aniso}$ & \boldmath$\sqrt{\rho_{\min}}$ & \boldmath$\sqrt{\rho_{\max}}$ \\
\hline
$\Lambda$CDM
& 3 & $0.070\pm0.026$ & $0.994\pm0.015$
& $0.038\pm0.019$ & $0.970\pm0.018$ & $1.016\pm0.021$ \\
$w$CDM
& 4 & $0.338\pm0.023$ & $1.004\pm0.011$
& $0.057\pm0.015$ & $0.962\pm0.015$ & $1.039\pm0.017$ \\
CPL
& 5 & $0.306\pm0.019$ & $0.998\pm0.009$
& $0.053\pm0.012$ & $0.966\pm0.012$ & $1.035\pm0.017$ \\
$\Lambda_{\rm s}$CDM
& 4 & $0.074\pm0.021$ & $1.005\pm0.010$
& $0.029\pm0.015$ & $0.994\pm0.013$ & $1.030\pm0.018$ \\
IDE
& 4 & $1.163\pm0.013$ & $0.573\pm0.007$
& $1.870\pm0.029$ & $0.114\pm0.004$ & $1.055\pm0.017$ \\
DCDMDR
& 4 & $0.977\pm0.018$ & $0.737\pm0.012$
& $1.380\pm0.027$ & $0.224\pm0.007$ & $1.212\pm0.027$ \\
$\Lambda w$DM
& 4 & $1.372\pm0.030$ & $1.404\pm0.017$
& $1.476\pm0.022$ & $0.738\pm0.016$ & $4.924\pm0.116$ \\
$\Lambda w(a)$DM
& 4 & $2.119\pm0.042$ & $0.828\pm0.011$
& $1.488\pm0.024$ & $0.246\pm0.006$ & $1.873\pm0.047$ \\
\hline
\end{tabular}
\caption{\texttt{PostModes} comparison between $A=$ full Planck-CMB+DESI+PantheonPlus and $B=$ compressed Planck-CMB+DESI+PantheonPlus. Here, $N$ denotes the dimensionality of the posterior subspace used for the comparison, comprising the three CMB-compression parameters together with the additional parameter(s) specific to each cosmological model. $S_{\rm CC}$ measures the displacement between the posterior centers in the covariance-combined metric, $\sqrt{\alpha}$ gives the geometric-mean rescaling of the posterior standard deviations, and $A_{\rm aniso}$ quantifies how strongly this rescaling varies across different parameter-space directions. The quantities $\sqrt{\rho_{\min}}$ and $\sqrt{\rho_{\max}}$ give, respectively, the smallest and largest ratios of the standard deviations of $B$ to those of $A$ along the generalized eigenmodes. Central values are obtained from the full chains, while quoted uncertainties are Monte Carlo standard errors estimated from 500 circular moving-block bootstrap replicates, retaining the sample weights and resampling each post-burn-in chain separately.}
\label{tab:postmodes_summary}
\end{table*}

The multidimensional diagnostics show that $\Lambda w(a)$DM exhibits the largest displacement between the posterior centers among the models considered, with $S_{\rm CC}=2.119$. At the same time, $\sqrt{\alpha}=0.828$ indicates an overall reduction of the geometric-mean posterior width in $B$. This average, however, hides a strongly anisotropic deformation, as shown by $A_{\rm aniso}=1.488$. The most contracted generalized mode has $\sqrt{\rho_{\min}}=0.246$, corresponding to a standard deviation approximately $4.1$ times smaller in $B$ than in $A$. In contrast, the broadest mode has $\sqrt{\rho_{\max}}=1.873$ and is therefore almost a factor of two broader in the compressed posterior. The difference between the two analyses is thus not a simple tightening or broadening of the same posterior, but a substantial redistribution of posterior widths across different parameter combinations.

\begin{figure*}[ht!]
    \centering
    \includegraphics[width=0.9\textwidth]{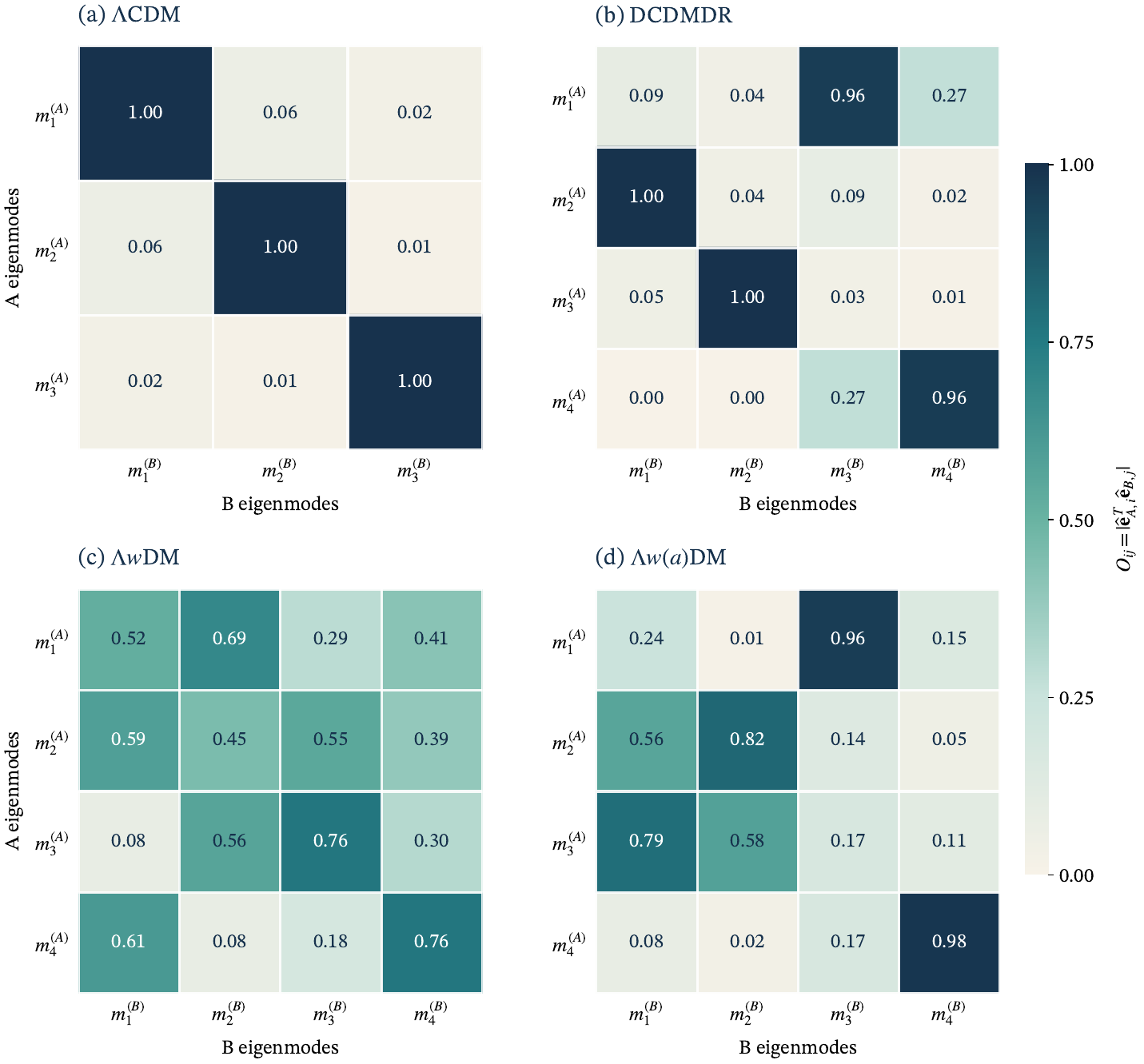}
    \caption{Eigenvector-overlap matrices for the reference posterior ($A=$ full Planck-CMB+DESI+PantheonPlus) and the alternative posterior ($B=$ compressed Planck-CMB+DESI+PantheonPlus). Each element is defined according to Eq.~\eqref{eq:overlap}, where the rows and columns correspond to the principal modes $m_i^{(A)}$ and $m_j^{(B)}$, respectively, ordered by decreasing eigenvalue of the corresponding standardized covariance matrix. The overlaps are evaluated in the common reference-standardized metric. $O_{ij}=1$ indicates aligned directions up to an arbitrary eigenvector sign, whereas $O_{ij}=0$ indicates orthogonal directions. The nearly diagonal $\Lambda$CDM matrix (top-left panel) shows that the principal basis is preserved. The off-diagonal structure in DCDMDR (top-right panel) is predominantly associated with a reordering of closely aligned modes and modest residual mixing. Both the constant-EoS $\Lambda w$DM model (bottom-left panel) and the time-dependent $\Lambda w(a)$DM model (bottom-right panel) exhibit substantial mixing among several modes.}
    \label{fig:eigenvector_overlap}
\end{figure*}

This is also evident from the eigenvector-overlap matrix shown in the bottom-right panel of Figure~\ref{fig:eigenvector_overlap}. In contrast to the nearly diagonal $\Lambda$CDM case, or the predominantly reordered modes observed in DCDMDR, here several principal directions mix substantially between $A$ and $B$. Even after accounting for the dominant correspondence between modes, some matched overlaps remain only at the level of $\sim0.8$. Geometrically, this means that, in addition to the large changes in directional widths, the principal axes themselves are rotated. Therefore, the parameter combinations defining the dominant degeneracy directions differ between the two posteriors.

\subsection{Overall pattern}
\label{sec:overall_pattern}

The results above reveal a clear separation in the performance of the compressed likelihood across the models considered. For $\Lambda$CDM, $w$CDM, CPL, and $\Lambda_{\rm s}$CDM, the compressed likelihood reproduces both the marginalized constraints and the multidimensional posterior geometry with high accuracy. In all four cases, the multidimensional diagnostics remain close to their reference values, with $\sqrt{\alpha}$ differing from unity by less than $\sim1\%$, $A_{\rm aniso}\lesssim 0.06$, and the extrema of $\sqrt{\rho_i}$ staying within $\sim4\%$ of unity.

IDE, DCDMDR, $\Lambda w$DM, and $\Lambda w(a)$DM instead show substantially larger differences between $A$ and $B$. Their center separations increase to $S_{\rm CC}\simeq0.98$--$2.12$, while $A_{\rm aniso}\simeq1.38$--$1.87$. In IDE, DCDMDR, and $\Lambda w(a)$DM, at least one generalized mode is substantially narrower in the compressed posterior, with $\sqrt{\rho_{\min}}\simeq0.11$--$0.25$, while other directions are much less affected or can become broader. In the constant-EoS $\Lambda w$DM model, the compressed posterior instead broadens overall, with $\sqrt{\alpha}=1.404$, and its broadest generalized mode has $\sqrt{\rho_{\max}}=4.924$. These geometric differences are reflected in substantial changes in the marginalized constraints and correlations involving the additional model parameters.

The origin of this systematic pattern and its implications for the validity of CMB compression are discussed in the following section.

%=======================================
\section{Discussion}
\label{sec:discussion}

\begin{figure*}[ht!]
    \centering
    \includegraphics[width=0.9\textwidth]{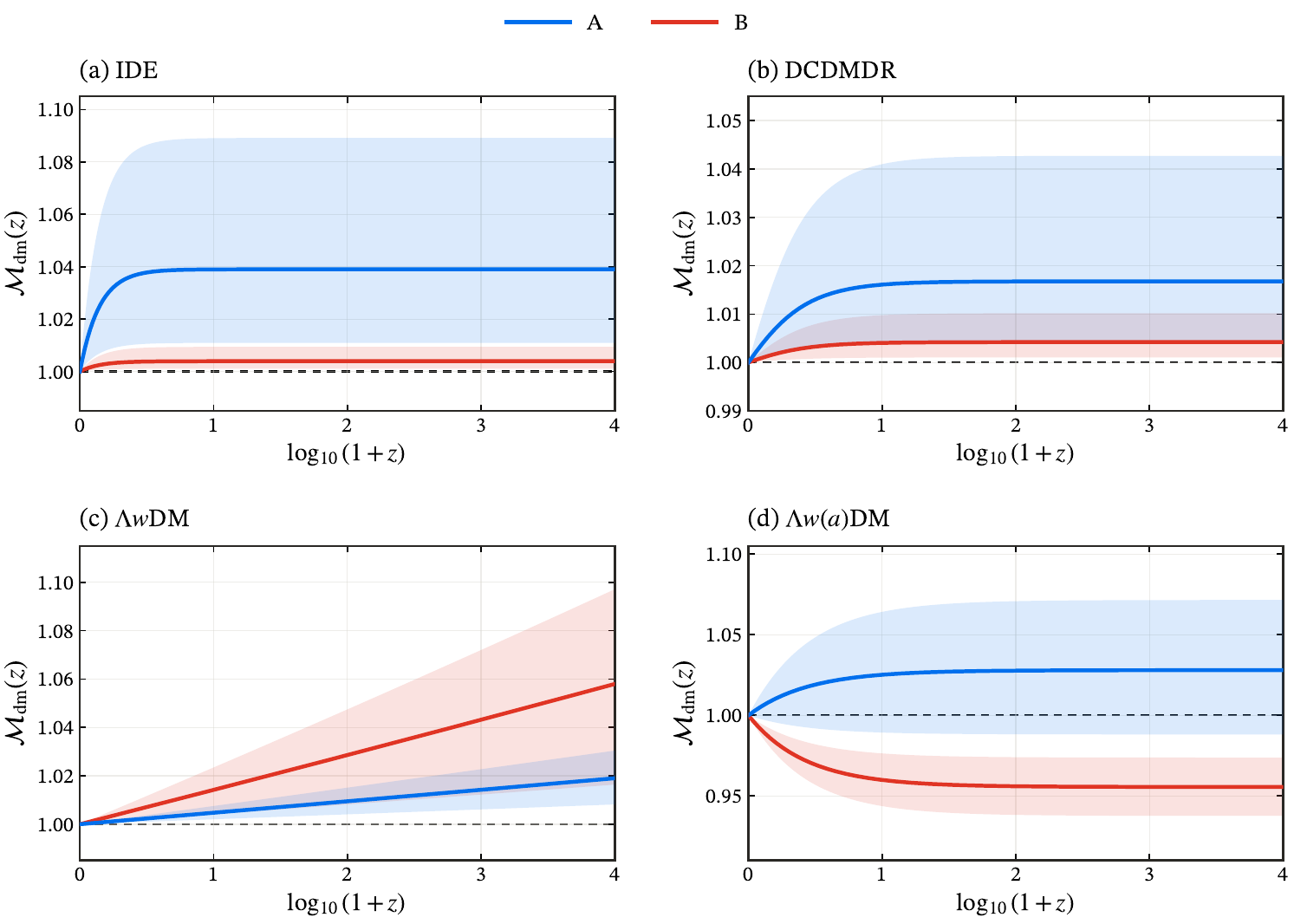}
    \caption{Evolution of the dark-matter mapping function ${\cal M}_{\rm dm}(z)\equiv \rho_{\rm dm}(z)/[\rho_{{\rm dm},0}(1+z)^3]$ for IDE (top left), DCDMDR (top right), $\Lambda w$DM (bottom left), and $\Lambda w(a)$DM (bottom right)}. Blue and red curves correspond to the posterior distributions obtained from $A$ and $B$, respectively. The shaded regions show the $68\%$ credible intervals. The horizontal dashed line ${\cal M}_{\rm dm}=1$ marks the standard pressureless dark matter relation.
    \label{fig:dm_mapping}
\end{figure*}

The results discussed in the previous section reveal a clear separation in the performance of the compressed CMB likelihood. In $\Lambda$CDM, $w$CDM, CPL, and $\Lambda_{\rm s}$CDM, the two prescriptions yield nearly identical marginalized constraints and multidimensional posterior geometries. Conversely, substantial differences emerge for IDE, DCDMDR, $\Lambda w$DM, and $\Lambda w(a)$DM. Since models introducing new physics that acts predominantly at late times appear on both sides of this separation, a simple distinction between early- and late-time modifications is not sufficient to determine whether the compression scheme remains reliable.

A more informative distinction is that all four models showing a breakdown directly modify the evolution of the dark-matter sector. Whenever the standard evolution of the matter components is preserved (i.e., $\rho_b\propto a^{-3}$ and $\rho_{\rm dm}\propto a^{-3}$), the present-day physical matter density $\omega_m=\omega_b+\omega_{\rm dm}+\omega_\nu$ is a natural compression variable because, for a fixed neutrino sector, it uniquely determines the corresponding matter abundance around recombination. This one-to-one mapping is preserved in $\Lambda$CDM, $w$CDM, CPL, and $\Lambda_{\rm s}$CDM. By contrast, in IDE, DCDMDR, $\Lambda w$DM, and $\Lambda w(a)$DM, this mapping is broken because the dark-matter density evolves non-standardly. In these models, fixing the same present-day value of $\omega_m$ implies a different dark-matter abundance (and hence a different total matter abundance) at recombination. A different matter abundance at recombination, in turn, can substantially modify the CMB angular power spectra by shifting the epoch of matter-radiation equality and changing the depth and evolution of the gravitational potentials that drive the acoustic oscillations~\cite{Eisenstein:2004an,Hu:2004kn,Baryakhtar:2024rky,Loverde:2024nfi}. Consequently, the same present-day value of $\omega_m$ no longer corresponds to the same physical conditions at recombination as in models that preserve the standard evolution of the matter components.

To test this interpretation directly, we proceed in three steps. We first quantify how strongly the standard mapping between the present-day and earlier dark-matter abundances is violated in the four models for which the compression fails. We then examine how this altered mapping changes the dark-matter abundance relevant at recombination and how this is reflected in the full and compressed CMB posteriors. Finally, we discuss the main caveats surrounding perturbative effects that are not captured by the background mapping alone.

\subsubsection{Quantifying the breakdown of the dark-matter scaling}

To quantify departures from the standard evolution of the dark-matter component, we define
\begin{equation}
{\cal M}_{\rm dm}(z)
\equiv
\frac{\rho_{\rm dm}(z)}
{\rho_{{\rm dm},0}(1+z)^3}.
\label{eq:dm_mapping}
\end{equation}
For DCDMDR, we apply the same definition to the decaying dark-matter component, replacing $\rho_{\rm dm}$ and $\rho_{{\rm dm},0}$ with $\rho_{\rm dcdm}$ and $\rho_{{\rm dcdm},0}$, respectively. For stable pressureless dark matter, ${\cal M}_{\rm dm}(z)=1$ at all redshifts. More generally, this quantity provides two complementary pieces of information. First, if the dark-matter density scales as $\rho_{\rm dm}\propto(1+z)^3$, then ${\cal M}_{\rm dm}(z)$ is constant. Its redshift dependence therefore directly signals departures from the standard dark-matter redshift evolution. Second, its absolute value quantifies the cumulative mismatch between the actual dark-matter abundance at a given redshift and that obtained by extrapolating the present-day abundance using the standard relation $\rho_{\rm dm}\propto(1+z)^3$. In particular, ${\cal M}_{\rm dm}>1$ means that the standard extrapolation underestimates the actual dark-matter abundance at a given redshift, whereas ${\cal M}_{\rm dm}<1$ means that it overestimates it.

We reconstruct ${\cal M}_{\rm dm}(z)$ from the posterior samples of $A$ and $B$ for the four models in which the compression shows a significant breakdown, over a redshift range extending to the epochs relevant to the CMB. The resulting central curves and $68\%$ credible regions are shown in Figure~\ref{fig:dm_mapping}.

For IDE, the evolution of ${\cal M}_{\rm dm}(z)$ follows directly from the conservation equation. At late times, the interaction modifies the relation between the present-day dark-matter density and its value in the past, see Eq.~\eqref{eq:IDE_rhodm}. At sufficiently high redshift, however, $\rho_{\rm DE}/\rho_{\rm dm}$ becomes small and the interaction term becomes negligible relative to the standard dilution term. The dark matter therefore approaches the standard scaling $\rho_{\rm dm}\propto(1+z)^3$, causing ${\cal M}_{\rm dm}(z)$ to approach a constant value, as seen in Figure~\ref{fig:dm_mapping}. Importantly, this asymptotic value is not equal to unity. Thus, although the dark matter recovers the standard redshift scaling at high redshift, its abundance remains offset from the standard backward extrapolation of the present-day density. In particular, extrapolating the present-day density backward according to that scaling would underestimate the actual dark-matter abundance at recombination.

As shown in the top-right panel of Figure~\ref{fig:dm_mapping}, DCDMDR exhibits the same qualitative behavior of ${\cal M}_{\rm dcdm}(z)$, but for a different physical reason. The density of the decaying dark-matter component evolves according to Eq.~\eqref{eq:DDM_rho_evolution}. Going backward in time, the expansion rate becomes increasingly large relative to the fixed physical decay rate, so that the fractional effect of the decay over a Hubble time becomes progressively smaller. The decaying dark matter therefore approaches the standard scaling $\rho_{\rm dcdm}\propto(1+z)^3$, causing ${\cal M}_{\rm dcdm}(z)$ to approach a constant value. As in IDE, however, this asymptotic value is not equal to unity. Extrapolating the present-day density backward according to the standard scaling would therefore underestimate the actual abundance of the decaying dark-matter component at recombination.

The behavior is qualitatively different for $\Lambda w$DM, as shown in the bottom-left panel of Figure~\ref{fig:dm_mapping}. In this case, the dark-matter density evolves according to Eq.~\eqref{eq:lwdm_background}, and therefore ${\cal M}_{\rm dm}(z)=(1+z)^{3w_{\rm dm,0}}$. Unlike IDE and DCDMDR, as long as $w_{\rm dm,0}\neq0$, the dark-matter density does not recover the standard $(1+z)^3$ scaling and ${\cal M}_{\rm dm}(z)$ does not approach a nonzero constant value at high redshift. Instead, the departure from the standard mapping grows continuously with redshift. The direction of this departure depends on the sign of the dark-matter EoS: for $w_{\rm dm,0}>0$, ${\cal M}_{\rm dm}>1$ and the standard backward extrapolation underestimates the actual dark-matter abundance, whereas for $w_{\rm dm,0}<0$, ${\cal M}_{\rm dm}<1$ and it overestimates it. In either case, the mismatch increases with redshift and can become substantial at the characteristic redshifts probed by the CMB. The persistent modification of the scaling means that fixing the dark-matter abundance at one epoch does not fix its evolution across the redshift range probed by the CMB.

The time-dependent $\Lambda w(a)$DM model instead recovers the standard scaling at high redshift, as shown in the bottom-right panel of Figure~\ref{fig:dm_mapping}. From Eq.~\eqref{eq:lwadm_background}, we see that the mapping ${\cal M}_{\rm dm}(z)=\exp[3w_{\rm dm,0}z/(1+z)]$ approaches $e^{3w_{\rm dm,0}}$ as $z\rightarrow\infty$. As in IDE and DCDMDR, the early-time departure is therefore predominantly a normalization offset. Its sign is set by $w_{\rm dm,0}$: a positive EoS gives ${\cal M}_{\rm dm}>1$, while a negative EoS gives ${\cal M}_{\rm dm}<1$. The different signs preferred by $A$ and $B$ consequently place their reconstructed mapping functions on opposite sides.

\subsubsection{Implications for the recombination-era dark-matter abundance}

\begin{figure*}[ht!]
    \centering
    \includegraphics[width=0.9\textwidth]{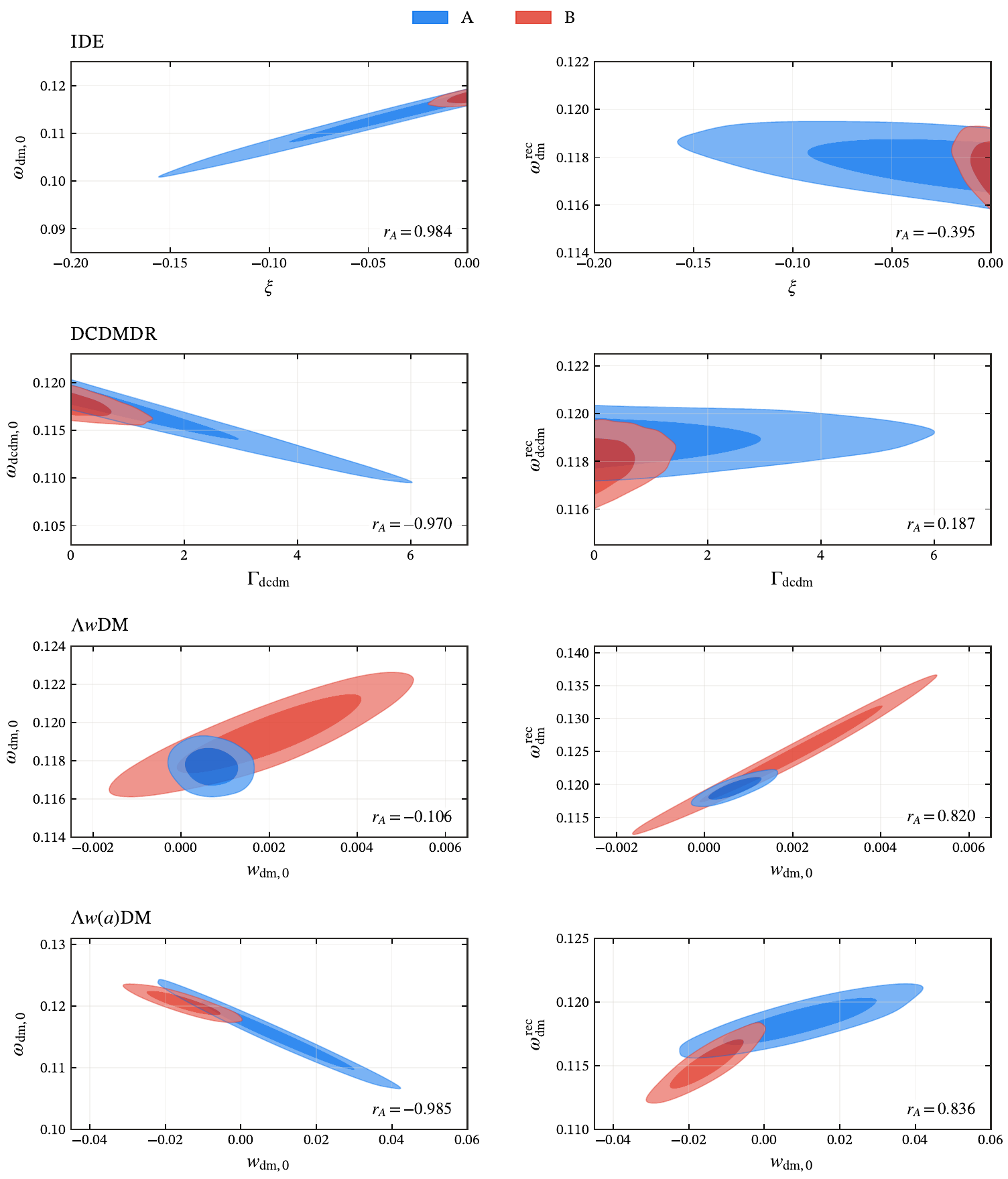}
    \caption{Comparison of the posterior distributions obtained using the present-day dark-matter density (left column) and the corresponding recombination-equivalent density defined in Eq.~\eqref{eq:omega_dm_rec} (right column), for IDE (first row), DCDMDR (second row), $\Lambda w$DM (third row), and $\Lambda w(a)$DM (fourth row)}. The values reported in each panel give the correlation coefficient $r_A$ computed from the full-CMB posterior $A$.
    \label{fig:dm_reparameterization}
\end{figure*}

To test more directly the role of the altered relation between the present-day and recombination-era dark-matter abundances, we define
\begin{equation}
\omega_{\rm dm}^{\rm rec}
\equiv
\omega_{\rm dm}\,
{\cal M}_{\rm dm}(z_{\rm rec}),
\label{eq:omega_dm_rec}
\end{equation}
where, as usual, $\omega_{\rm dm}$ denotes the present-day physical dark-matter density and $z_{\rm rec}\simeq1100$. The quantity $\omega_{\rm dm}^{\rm rec}$ is the equivalent present-day physical density that, if evolved backward according to the standard $\rho_{\rm dm}\propto(1+z)^3$ scaling, would reproduce the actual dark-matter abundance of the model at recombination. It therefore provides a direct diagnostic of the altered relation between the two epochs: if the standard evolution is preserved, ${\cal M}_{\rm dm}(z_{\rm rec})=1$ and $\omega_{\rm dm}^{\rm rec}=\omega_{\rm dm}$, whereas any difference between the two quantities directly reflects the non-standard dark-matter evolution accumulated between today and recombination. 

Figure~\ref{fig:dm_reparameterization} compares, for each of the four models, the correlations between the additional model parameter and the dark-matter density before and after the reparameterization. In the left panels, we show the two-dimensional posterior distributions involving the present-day dark-matter density, $\omega_{\rm dm}$, while in the right panels we show the corresponding distributions after replacing it with the recombination-equivalent density, $\omega_{\rm dm}^{\rm rec}$. 

To understand the physical origin of these correlations, we first focus on the full-CMB posterior $A$. For IDE, the original parameterization exhibits an almost one-dimensional degeneracy between $\xi$ and $\omega_{\rm dm}$, as shown in the top-left panel of Figure~\ref{fig:dm_reparameterization}. For the negative coupling considered here, a stronger interaction transfers more energy from dark matter to dark energy between recombination and today, so that a more negative $\xi$ requires a smaller present-day $\omega_{\rm dm}$ to preserve the same dark-matter abundance around recombination. This generates the strong $\xi$--$\omega_{\rm dm}$ degeneracy along which the resulting CMB spectra remain similar. This compensating direction is largely lost in the compressed posterior $B$. The compressed likelihood directly constrains $\omega_m$ according to its $\Lambda$CDM calibration. It therefore penalizes the decrease in $\omega_{\rm dm}$ required by a more negative $\xi$, even though this combination preserves the same dark-matter abundance at recombination. The loss of this degeneracy artificially reduces the allowed parameter space and leads to spuriously tighter constraints. The reparameterization in terms of $\omega_{\rm dm}^{\rm rec}$ tests this explanation directly. As shown in the top-right panel of Figure~\ref{fig:dm_reparameterization}, the correlation changes from $r_A(\xi,\omega_{\rm dm})=0.984$ to $r_A(\xi,\omega_{\rm dm}^{\rm rec})=-0.395$.\footnote{For two parameters $x$ and $y$, the correlation coefficient is defined as $r(x,y)\equiv{\rm Cov}(x,y)/[\sigma(x)\sigma(y)]$.} Since $\omega_{\rm dm}^{\rm rec}$ includes the effect of the non-standard dark-matter evolution between today and recombination, different combinations of $\xi$ and $\omega_{\rm dm}$ that produce the same dark-matter abundance at recombination correspond to the same $\omega_{\rm dm}^{\rm rec}$. The reduced correlation therefore confirms that most of the original $\xi$--$\omega_{\rm dm}$ degeneracy originates from the altered mapping between the two epochs. This interpretation is also consistent with the multidimensional results of Table~\ref{tab:postmodes_summary}, where the breakdown in IDE appears predominantly as a strongly anisotropic deformation of the posterior rather than as a uniform change in its scale.

A very similar picture emerges for DCDMDR, see the second row of Figure~\ref{fig:dm_reparameterization}. The strong anticorrelation between the decay rate and the present-day density of the decaying dark-matter component, $r_A(\Gamma_{\rm dcdm},\omega_{\rm dcdm})=-0.970$, is reduced to $r_A(\Gamma_{\rm dcdm},\omega_{\rm dcdm}^{\rm rec})=0.187$ when the density is expressed in terms of $\omega_{\rm dcdm}^{\rm rec}$. For a fixed abundance today, a larger decay rate implies that more dark matter must have been present at recombination. Conversely, preserving the same recombination-era abundance requires a smaller present-day surviving dark-matter density as the decay rate increases. The strong anticorrelation in the original parameterization therefore reflects this compensation and is largely removed when the posterior is expressed in terms of $\omega_{\rm dcdm}^{\rm rec}$. This interpretation is also consistent with the strongly anisotropic posterior deformation summarized in Table~\ref{tab:postmodes_summary}. In DCDMDR, however, varying the decay rate also changes the redshift-dependent partition of the dark sector between the decaying matter component and the dark radiation produced by the decay. This additional redistribution may contribute to the more complex changes in the posterior geometry, including the reordering of principal modes observed in the top-right panel of Figure~\ref{fig:eigenvector_overlap}.

The situation is different for the constant-EoS $\Lambda w$DM model, shown in the third row of Figure~\ref{fig:dm_reparameterization}. The full-CMB posterior has only a weak correlation between $w_{\rm dm,0}$ and the present-day density, $r_A(w_{\rm dm,0},\omega_{\rm dm})=-0.106$, while a strong positive correlation appears after the reparameterization, $r_A(w_{\rm dm,0},\omega_{\rm dm}^{\rm rec})=0.820$. Unlike IDE and DCDMDR, the full-CMB constraint therefore does not follow a strong compensating degeneracy between the additional parameter and the present-day dark-matter density. Moreover, specifying $\omega_{\rm dm}^{\rm rec}$ fixes the abundance at only one epoch, whereas $w_{\rm dm,0}$ changes its scaling across all redshifts. Defining the effective present-day density corresponding to the abundance at an arbitrary redshift $z$, we obtain
\begin{equation}
\omega_{\rm dm}^{\rm eff}(z)
=
\omega_{\rm dm}^{\rm rec}
\left(
\frac{1+z}{1+z_{\rm rec}}
\right)^{3w_{\rm dm,0}}.
\label{eq:lwdm_history}
\end{equation}
At recombination, $\omega_{\rm dm}^{\rm eff}(z_{\rm rec})=\omega_{\rm dm}^{\rm rec}$ by construction. At any other redshift, however, $\omega_{\rm dm}^{\rm eff}(z)$ retains an explicit dependence on $w_{\rm dm,0}$. Two models with the same $\omega_{\rm dm}^{\rm rec}$ but different $w_{\rm dm,0}$ therefore agree on the dark-matter abundance at recombination but predict different abundances at matter-radiation equality, at the drag epoch, and throughout the redshift range relevant to the primary CMB anisotropies. The effect of $w_{\rm dm,0}$ cannot therefore be absorbed into a single recombination-equivalent density. The full CMB can constrain this evolution through changes in the acoustic structure and gravitational potentials that are not specified by the three compressed variables. The tight full-CMB constraint on $w_{\rm dm,0}$, together with its weak correlation with the present-day density, is consistent with sensitivity to information beyond a single early-time abundance.

For the time-dependent $\Lambda w(a)$DM model, the fourth row of Figure~\ref{fig:dm_reparameterization} shows a strong anticorrelation between $w_{\rm dm,0}$ and the present-day density, $r_A(w_{\rm dm,0},\omega_{\rm dm})=-0.985$. This is because increasing $w_{\rm dm,0}$ raises the early-time normalization of the dark-matter density, which can be compensated by lowering its present-day value to preserve the same early-time abundance. Expressing the density in terms of $\omega_{\rm dm}^{\rm rec}$ reverses the sign of the correlation, giving $r_A(w_{\rm dm,0},\omega_{\rm dm}^{\rm rec})=0.836$. Thus, although the reparameterization accounts for the altered mapping between the two epochs, it does not eliminate the correlation between the EoS parameter and the reconstructed abundance. The effective density at an arbitrary redshift can be written as
\begin{equation}
\omega_{\rm dm}^{\rm eff}(z)
=
\omega_{\rm dm}^{\rm rec}
\exp\left[3w_{\rm dm,0}
\left(\frac{1}{1+z_{\rm rec}}-\frac{1}{1+z}\right)\right].
\label{eq:lwadm_history}
\end{equation}
Unlike Eq.~\eqref{eq:lwdm_history}, at fixed $\omega_{\rm dm}^{\rm rec}$ the residual dependence on $w_{\rm dm,0}$ is suppressed when both $z$ and $z_{\rm rec}$ are large.

The opposite changes in constraining power can be understood in terms of which information is distorted or lost by the compression. In IDE, DCDMDR, and $\Lambda w(a)$DM, the full-CMB posterior allows a strong compensation between the additional model parameter and the present-day dark-matter density. The $\Lambda$CDM-calibrated constraint on $\omega_m$ restricts this degeneracy and can therefore produce artificially tighter marginalized constraints. In the constant-EoS $\Lambda w$DM model, instead, the departure from standard dark-matter evolution persists throughout the epochs relevant to the primary anisotropies. The full spectra can constrain this evolution beyond its effect on a single early-time abundance, whereas the three compressed quantities do not capture that information. This provides an interpretation of the uncertainty on $w_{\rm dm,0}$ being larger in $B$ by a factor of $\sim3.5$.

\subsubsection{Caveats surrounding cosmological perturbations}

The background-level mapping identified above does not exhaust the information carried by the full CMB likelihood. Importantly, IDE, DCDMDR, $\Lambda w$DM, and $\Lambda w(a)$DM also modify the evolution of cosmological perturbations, which can introduce further model-dependent changes in the CMB spectra and in the corresponding parameter degeneracies. The role of perturbations should therefore be distinguished according to three different cases.
\begin{itemize}
\item If a model modifies both the dark-matter background evolution and its perturbations, as in IDE, DCDMDR, $\Lambda w$DM, and $\Lambda w(a)$DM, the breakdown of the background mapping discussed here is sufficient to invalidate the standard compressed likelihood. Perturbative effects can add further model-dependent information to the CMB anisotropies and posterior degeneracies that is not retained by the compression. However, they cannot restore the background relation on which the compression is based.
\item If the standard dark-matter background evolution is preserved and the background modification is confined to late-time dark energy, the situation is different. For the models considered here, namely $w$CDM, CPL, and $\Lambda_{\rm s}$CDM, we find that the compression reproduces the full-CMB constraints well. This result should nevertheless be interpreted with the caveat that more general dark-energy models may exhibit significant perturbative effects that are not captured by the geometric compression. For example, a sufficiently large modification of the late-time ISW contribution could introduce additional information even if the background mapping and acoustic-scale geometry remain well preserved. The perturbation sector should therefore be checked separately before extending the agreement found here to more general dark-energy models.
\item Finally, if the new physics modifies only cosmological perturbations while leaving the background unchanged, a CMB compression based only on background and acoustic-scale quantities is by construction insensitive to the relevant new information. In this case, the compressed likelihood cannot be expected to provide meaningful constraints on the additional physics.
\end{itemize}

\section{Conclusions}
\label{sec:conclusions}

In this work, we have investigated when compressed CMB likelihoods can be reliably used beyond $\Lambda$CDM, why they fail for some classes of models, and how such failures propagate into the inferred cosmological constraints. 

To address these questions, we considered a broad set of theoretical extensions that modify different sectors and epochs of the cosmological model. We considered late-time dark-energy modifications with a constant EoS ($w$CDM), smooth dynamical evolution (CPL), and an abrupt transition ($\Lambda_{\rm s}$CDM). We also considered models that modify the dark-matter evolution predominantly at late times, either through an interaction with dark energy (IDE) or through decay into dark radiation (DCDMDR). Finally, we considered two dark-matter EoS models: $\Lambda w$DM, with a constant $w_{\rm dm}(a)=w_{\rm dm,0}$ that modifies the redshift scaling throughout cosmic history, and $\Lambda w(a)$DM, with $w_{\rm dm}(a)=w_{\rm dm,0}a$ that approaches pressureless evolution at high redshift but changes the mapping between the present-day and early-time abundances.

For each model, we compared the constraints obtained from two datasets. Dataset $A$ combines the full Planck-CMB likelihood with DESI and PantheonPlus data. Dataset $B$ replaces the full CMB likelihood with the compressed CMB likelihood calibrated from the $\Lambda$CDM CMB-only analysis, keeping the DESI and PantheonPlus data unchanged. We compared both the marginalized parameter constraints and the full multidimensional posterior geometry using \texttt{PostModes}, a package developed in this work to quantify differences between high-dimensional posterior distributions.

We formulate concrete criteria for \textit{when}, \textit{why}, and \textit{how} the CMB compression fails.

\begin{itemize}
\item \textbf{When:} The relevant distinction is not whether the new physics acts at early or late times, but whether the model preserves the physical relations assumed by the compressed variables. Three cases should be distinguished. \textit{First}, if a model modifies the background evolution of dark matter, as in IDE, DCDMDR, $\Lambda w$DM, and $\Lambda w(a)$DM, the standard relation between the present-day matter density and the dark-matter abundance relevant to the CMB is no longer preserved. The background-level violation is already sufficient to invalidate the standard compressed likelihood. Perturbative effects may introduce additional model-dependent differences, but they cannot restore the background mapping on which the compression was calibrated. \textit{Second}, if the standard dark-matter evolution is preserved and the background modification is confined to the late-time dark-energy sector, the compression can remain reliable, as we find for $w$CDM, CPL, and $\Lambda_{\rm s}$CDM. Nevertheless, this result is not a general guarantee for arbitrary dark-energy models. Significant perturbative effects, for example, a large modification of the late-time ISW contribution, could carry CMB information that is not retained by the compressed likelihood and should therefore be checked case by case. \textit{Third}, if the new physics enters only through cosmological perturbations while leaving the background unchanged, a compression based on background and geometric quantities alone is by construction insensitive to the relevant new information and cannot be expected to constrain that physics.

\item \textbf{Why:} The CMB spectra, and therefore the full CMB likelihood, are primarily sensitive to the matter abundance at the epochs relevant to the formation of the primary CMB anisotropies, particularly matter-radiation equality and recombination. When the standard matter evolution is preserved, the present-day physical matter density uniquely determines the corresponding abundance at the redshifts probed by the CMB. This one-to-one relation is what makes $\omega_m$ a useful compression variable. If the dark-matter evolution is modified, however, this mapping can fail in two qualitatively different ways. First, the dark matter may recover the standard $(1+z)^3$ scaling at high redshift while retaining a different normalization, so that the present-day density still extrapolates to the wrong abundance around recombination. This happens in IDE, DCDMDR, and $\Lambda w(a)$DM, where the new physics modifies the dark-matter evolution predominantly at late times. Second, the redshift scaling itself may remain non-standard at all relevant redshifts. This occurs in $\Lambda w$DM, where the dark-matter EoS is allowed to deviate from zero and modifies its scaling throughout the cosmological evolution. In either case, preserving the matter abundance at a given early epoch requires adjusting the present-day dark-matter density $\omega_{\rm dm}$ accordingly. For fixed baryon and neutrino sectors, this necessarily shifts the present-day matter density $\omega_m$ as well. The full CMB likelihood evaluates the actual spectra for each parameter combination and can therefore accommodate such shifts when changes in the additional model parameter keep the CMB spectra nearly unchanged. The compressed likelihood instead constrains $\omega_m$ using the distribution calibrated in $\Lambda$CDM. It therefore penalizes shifts in the present-day dark-matter density through the corresponding change in $\omega_m$, even along degeneracy directions that leave the CMB spectra nearly unchanged. As a result, degeneracy directions present in the full CMB likelihood are lost or distorted in the compressed analysis. For the constant-EoS $\Lambda w$DM model, however, matching the abundance at one epoch does not preserve the evolution across the full redshift range probed by the CMB. The full spectra can constrain this persistent modification beyond its effect on a single early-time abundance. Losing part of this information in the compression can weaken the constraints instead of artificially tightening them.

\item \textbf{How:} An incorrect use of the CMB compression can deform the geometry of the multidimensional posterior and distort the marginalized cosmological constraints. We show that this can occur in several qualitatively different ways. The compression can shift the posterior center, compress or broaden different directions by very different amounts, alter the relative ordering of principal modes, and, in more severe cases, rotate and mix the eigendirections themselves. The models studied here provide concrete examples of these possibilities. In IDE, the dominant effect is a strongly anisotropic deformation associated with the altered relation between $\xi$ and $\omega_{\rm dm}$. In DCDMDR, the compression changes the relative widths of nearly preserved principal directions enough to reorder the modes, with the conversion of dark matter into dark radiation further complicating the posterior geometry. In $\Lambda w$DM and $\Lambda w(a)$DM, the failure extends beyond anisotropic changes in the posterior widths and produces genuine mixing and reorientation of the principal eigendirections. When these multidimensional distortions are projected onto the one- and two-dimensional marginalized constraints normally reported in cosmological analyses, they can appear as shifted central values or substantially tightened or weakened bounds. The resulting differences can be very large. For example, the constraint on $\xi$ in IDE differs by nearly an order of magnitude, while the upper bound on $\Gamma_{\rm dcdm}$ in DCDMDR changes by about a factor of four. The two EoS models illustrate the opposite changes in constraining power: the uncertainty on $w_{\rm dm,0}$ becomes larger by a factor of $\sim3.5$ in the constant-EoS case, but smaller by a factor of $\sim2$ in the time-dependent case when the full CMB is replaced by the compressed likelihood. It is also important to stress that agreement in a well-measured quantity such as the acoustic scale is not sufficient to demonstrate that the full posterior geometry has been preserved.
\end{itemize}

In conclusion, our results identify a restricted regime in which compressed CMB likelihoods can be used reliably: late-time dark-energy extensions that preserve the standard dark-matter background evolution and do not introduce significant additional perturbative effects. Outside this regime, the burden of proof should be reversed: the full CMB likelihood should be the default, and the compressed likelihood should be used only after its validity has been explicitly demonstrated for the model under consideration.

\begin{acknowledgments}
\noindent We thank Sunny Vagnozzi for several discussions. We acknowledge IT services at the University of Sheffield and at the University of Trento for the provision of services for High Performance Computing. D.P.\ acknowledges support from the Istituto Nazionale di Fisica Nucleare (INFN) through the Commissione Scientifica Nazionale 4 (CSN4) Iniziativa Specifica ``Quantum Fields in Gravity, Cosmology and Black Holes'' (FLAG).
W.G.\ acknowledges support from NASA under Grant No. 80NSSC24K0898. E.D.V.\ is supported by a Royal Society Dorothy Hodgkin Research Fellowship. This publication is based upon work from the COST Action CA21136 ``Addressing observational tensions in cosmology with systematics and fundamental physics'' (CosmoVerse), supported by COST (European Cooperation in Science and Technology).
\end{acknowledgments}
%%%%%%%%%%%%%%%%%%%%%%%%%%%%%%%%%%%%%%%%%%%%%%%%%%%%%%%%%
%%%%%%%%%%%%%%%%%%%%%%%%%%%%%%%%%%%%%%%%%%%%%%%%%%%%%%%%%

\appendix

\section{Robustness of the main results}
\label{Appendix:A}

\begin{figure*}[htpb!]
    \centering
    \includegraphics[width=0.85\columnwidth]{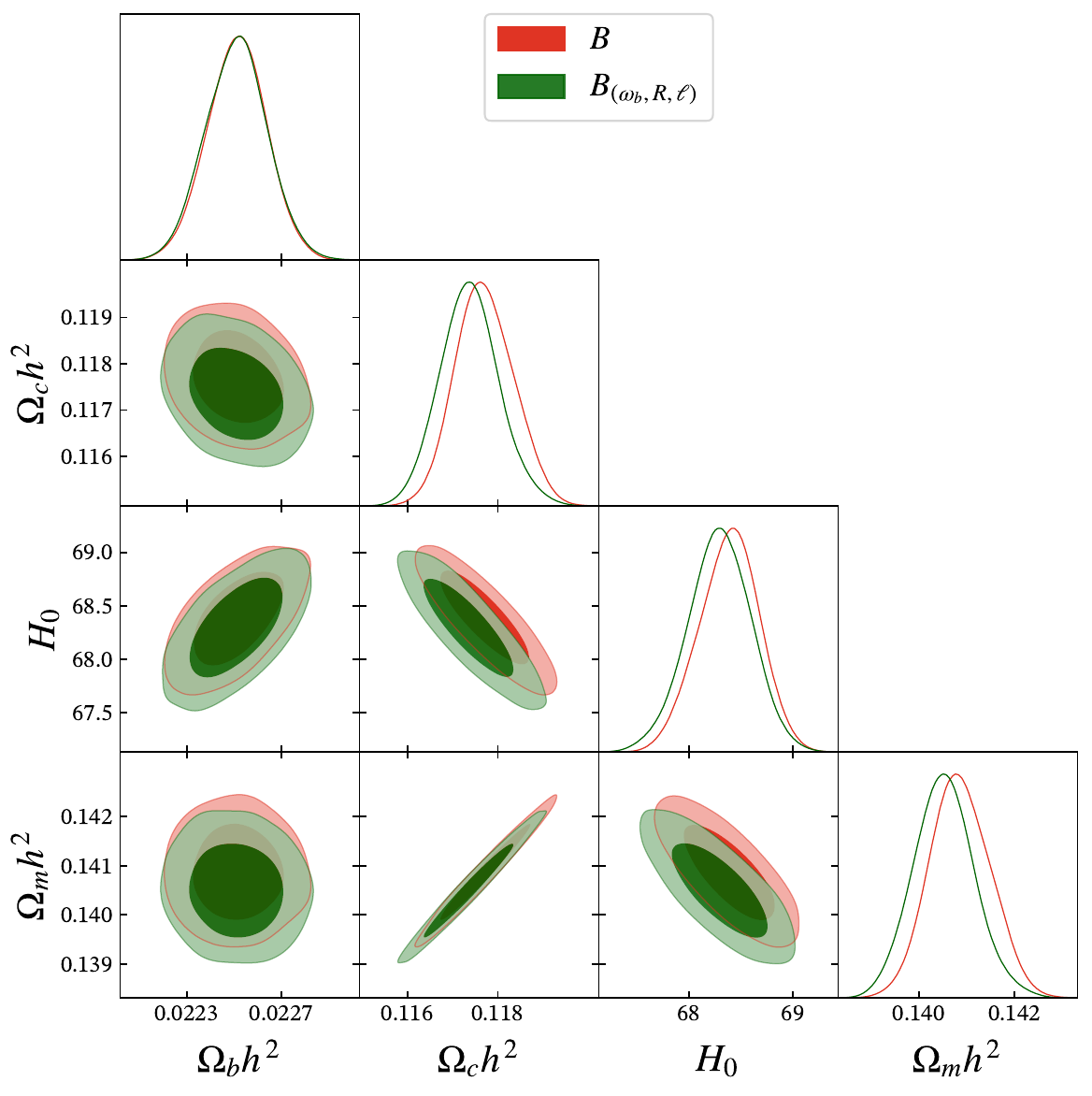}
    \includegraphics[width=0.85\columnwidth]{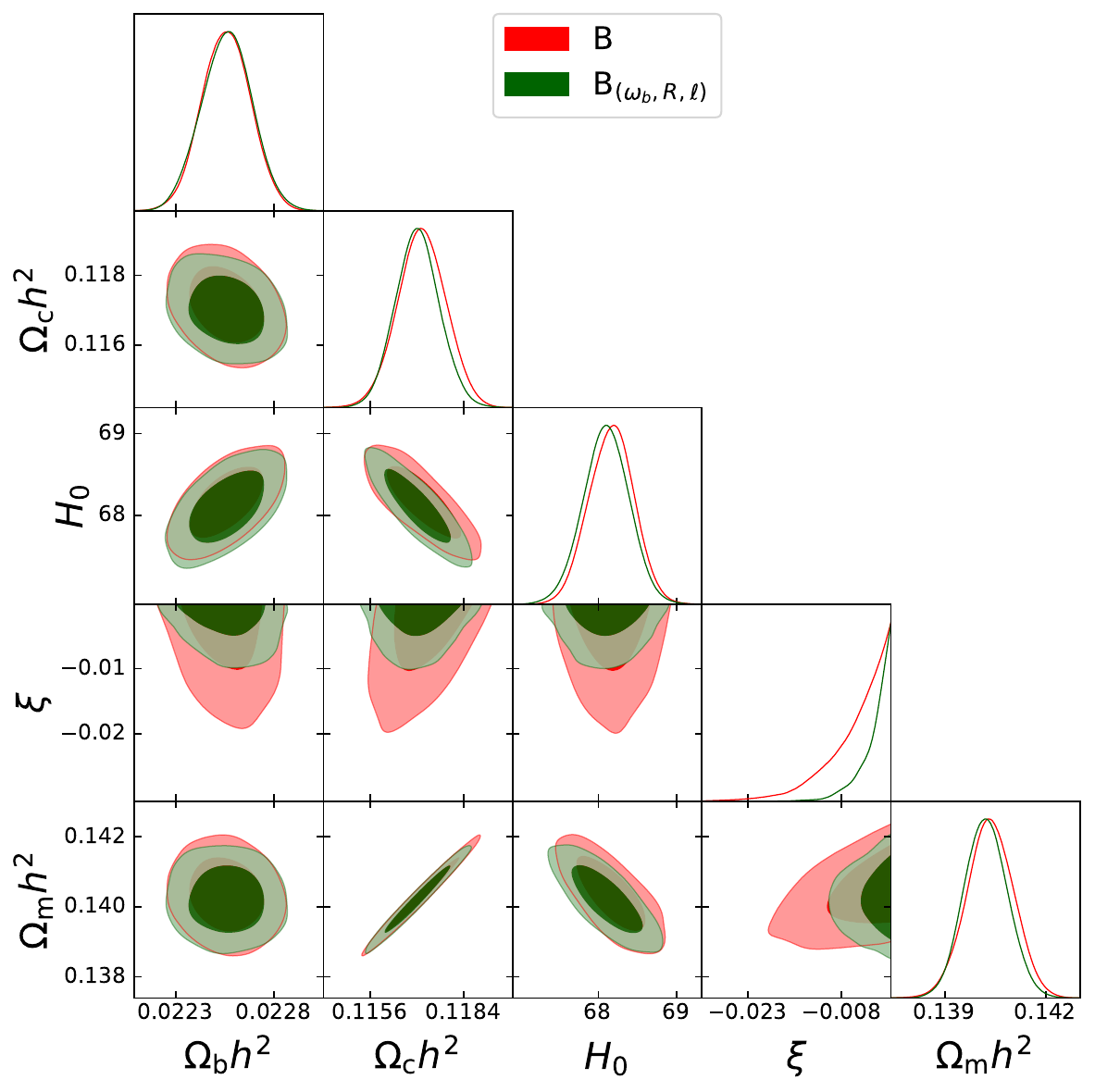}
    \caption{Comparison of the one- and two-dimensional posterior distributions obtained using the baseline compression basis $(1/\theta_s^\star,\omega_m,\omega_b)$ and the standard distance-prior basis $(R,\ell_a,\omega_b)$. The left panel shows $\Lambda$CDM. The right panel shows IDE.}
    \label{fig:test_WangWang}
\end{figure*}

The analysis presented in this work relies on specific choices for the parametrization of the compressed CMB likelihood, which we express in terms of the vector $\mathbf{x}=(1/\theta_s^\star,\omega_m,\omega_b)$, as well as for its combination with DESI BAO and PantheonPlus SN data. In this appendix, we show that the conclusions of the main text remain unchanged when the compression basis is replaced by the distance-prior variables $(R,\ell_a,\omega_b)$, when the dimensionality of the compressed likelihood is reduced to $\mathbf{x}=(1/\theta_s^\star,\omega_b)$, and when different combinations of low-redshift datasets are considered.

We have performed these robustness tests for all the models analyzed in this work. For conciseness, however, we report results only for $\Lambda$CDM, as a validation case, and IDE, as a representative example of the models for which the standard compression fails.

\subsection{Dependence on the compression basis}
\label{sec:WangWang}

We first test whether our results depend on the particular compression basis adopted in the main analysis. We compare our baseline parametrization, $(1/\theta_s^\star,\omega_m,\omega_b)$, with the standard distance-prior basis $(R,\ell_a,\omega_b)$. 

As a consistency check, the left panel of Figure~\ref{fig:test_WangWang} shows that the two compressed likelihoods give essentially identical constraints in $\Lambda$CDM, confirming that they encode the same CMB information when the standard matter evolution is preserved.

The right panel of Figure~\ref{fig:test_WangWang} shows the corresponding comparison for IDE. The two compression bases again produce similar posterior distributions. In particular, replacing $(1/\theta_s^\star,\omega_m,\omega_b)$ with $(R,\ell_a,\omega_b)$ does not restore the extended degeneracy allowed by the full CMB likelihood. The breakdown identified in the main text is therefore not a consequence of the specific compression basis adopted in our analysis, but of the physical mapping assumed by both parametrizations. We have verified the same behavior for all the models discussed in the main analysis.

\begin{table*}[htpb!]
\centering
\begin{tabular}{lccc}
\hline
Parameter & $A$ & $B$ & $C$ \\
\hline
$\Omega_{\mathrm b}h^2$ & $0.02249\pm0.00013$ & $0.02256\pm0.00012$ & $0.02234\pm 0.00015$ \\
$\Omega_{\mathrm c}h^2$ & $0.1125^{+0.0050}_{-0.0018}$ & $0.11714\pm0.00072$ & $0.103^{+0.018}_{-0.0097}$\\
$H_0$ & $68.48^{+0.34}_{-0.43}$ & $68.17\pm0.29$ & $65.4\pm 1.2$ \\
$\xi$ & $>-0.124$ & $>-0.0150$  & $> -0.444$\\
$\Omega_{\mathrm m}$ & $0.289^{+0.014}_{-0.0065}$ & $0.3021\pm0.0038$ & $0.293^{+0.035}_{-0.027}$\\
$100\theta_{\mathrm s}$ & $1.04208\pm0.00027$ & $1.04212\pm0.00028$ & $1.04185\pm 0.00029$ \\
\hline
\end{tabular}
\caption{Marginalized constraints on the most informative parameters of the IDE model. The datasets $A$ and $B$ are those defined in the main text, while $C$ is obtained by replacing the $3\times3$ compressed CMB likelihood with its $2\times2$ counterpart.}
\label{tab:app2}
\end{table*}

\begin{figure}[htpb!]
    \centering
    \includegraphics[width=0.9\linewidth]{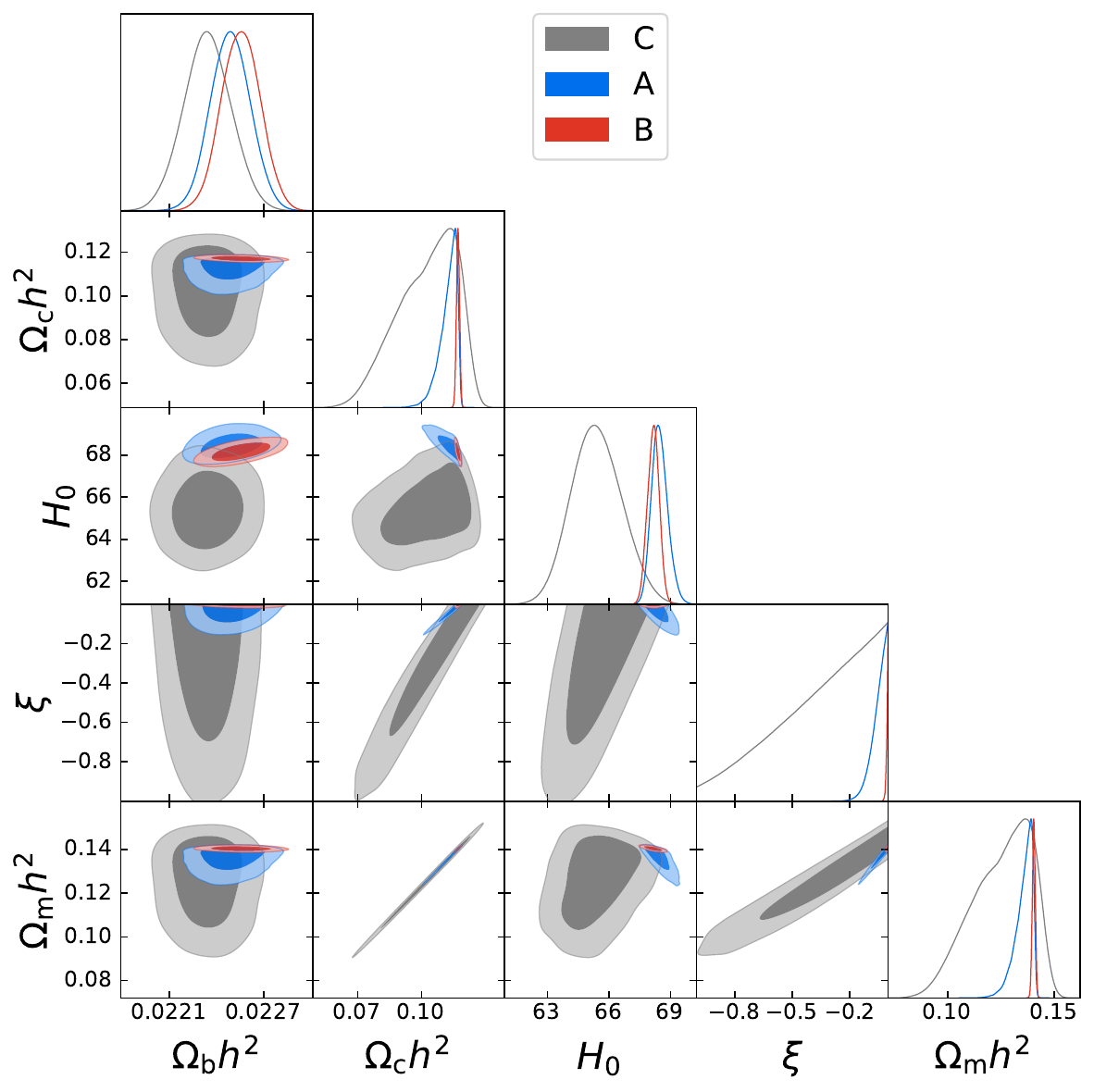}
    \caption{Comparison of the one- and two-dimensional posterior distributions in the IDE model. $A$ and $B$ are the datasets defined in the main text, while $C$ is obtained by replacing the $3\times3$ compressed CMB likelihood with its $2\times2$ counterpart.}
    \label{fig:app2}
\end{figure}

\subsection{Dependence on the compression dimensionality}

We next test whether the failure of the compressed likelihood can be avoided by reducing the dimensionality of the compression. To this end, we construct a two-dimensional compressed CMB likelihood retaining only $(1/\theta_s^\star,\omega_b)$. We denote by $C$ the dataset obtained by combining this $2\times2$ compressed CMB likelihood with DESI BAO and PantheonPlus SN. We compare it with datasets $A$ and $B$, defined as in the main analysis. Figure~\ref{fig:app2} and Table~\ref{tab:app2} therefore allow us to isolate the effect of reducing the dimensionality of the CMB compression.

Reducing the compression to two dimensions does not recover the full-CMB constraints. Instead, the nature of the disagreement changes. While the $3\times3$ compression strongly restricts the IDE degeneracy, removing $\omega_m$ discards a substantial amount of CMB information and produces much broader constraints. For example, the full-CMB analysis gives $\xi>-0.124$, compared with $\xi>-0.0150$ for the $3\times3$ compression and $\xi>-0.444$ for the $2\times2$ compression. 

The $2\times2$ compression also produces different correlations among the cosmological parameters, so that the much broader posterior does not simply encompass the full-CMB result. As visible in Figure~\ref{fig:app2}, the overlap with the full-CMB posterior remains limited in several projections despite the substantially larger uncertainties. The failure of the compressed likelihood is therefore not specific to the inclusion of $\omega_m$ as a third compressed parameter. Removing this information does not reconstruct the full likelihood, but instead leads to a different loss of CMB information and a different posterior geometry. 

We have verified the same behavior for all the models in which the full and compressed CMB likelihoods lead to discrepant constraints in the main analysis.

\subsection{Dependence on the dataset combination}

\begin{figure}[htpb!]
    \centering
    \includegraphics[width=\columnwidth]{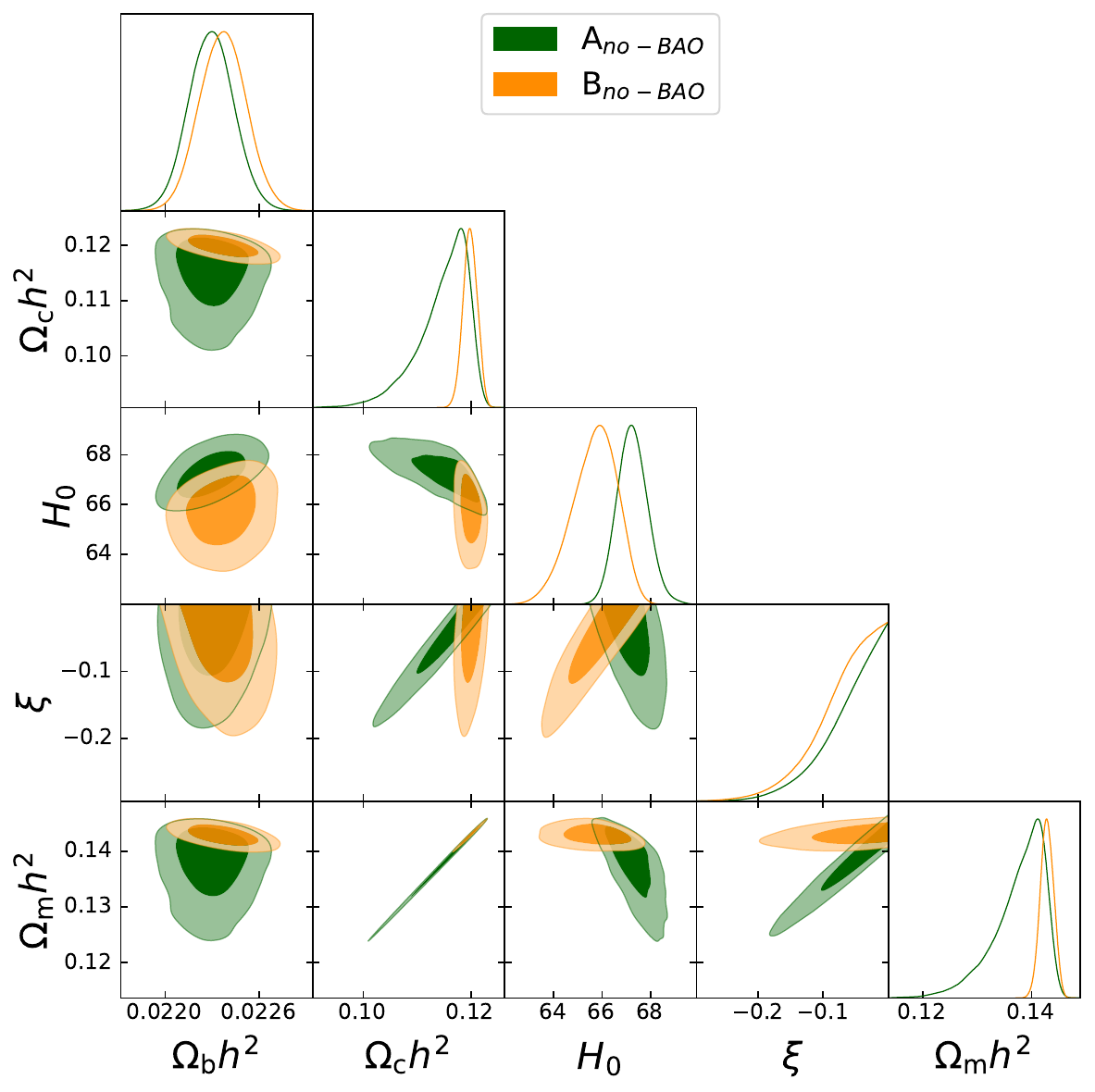}
    \caption{Comparison of the one- and two-dimensional posterior distributions for the IDE model using the $A_{\rm no\text{-}BAO}$ and $B_{\rm no\text{-}BAO}$ dataset combinations. These correspond to the datasets $A$ and $B$ defined in the main text after removing the DESI BAO likelihood.}
    \label{fig:app3}
\end{figure}

\begin{table*}[htpb!]
\centering
\begin{tabular}{l|cc|cc|cc}
\hline
Parameter & $A$ & $B$ & $A_{\rm no-SN}$ & $B_{\rm no-SN}$ & $A_{\rm no-BAO}$ & $B_{\rm no-BAO}$ \\
\hline
$\Omega_{\mathrm b}h^2$ & $0.02249\pm0.00025$ & $0.02256\pm0.00024$ & $0.02247^{+0.00026}_{-0.00025}$ & $0.02257\pm0.00025$ & $0.02229^{+0.00029}_{-0.00028}$ & $0.02237^{+0.00030}_{-0.00029}$ \\
$\Omega_{\mathrm c}h^2$ & $0.1125^{+0.0059}_{-0.0084}$ & $0.1171\pm0.0014$ & $0.0998^{+0.018}_{-0.022}$ & $0.1169\pm0.0015$ & $0.1148^{+0.0073}_{-0.0099}$ & $0.1198^{+0.0025}_{-0.0026}$ \\
$H_0$ & $68.48^{+0.81}_{-0.75}$ & $68.17\pm0.56$ & $69.4^{+1.6}_{-1.4}$ & $68.25^{+0.57}_{-0.58}$ & $67.3^{+1.3}_{-1.2}$ & $65.7^{+1.7}_{-1.8}$ \\
$\xi$ & $>-0.124$ & $>-0.0150$ & $>-0.330$ & $>-0.0164$ & $>-0.147$ & $>-0.156$ \\
$\Omega_{\mathrm m}$ & $0.289^{+0.019}_{-0.024}$ & $0.3021^{+0.0073}_{-0.0070}$ & $0.256^{+0.048}_{-0.056}$ & $0.3009^{+0.0076}_{-0.0073}$ & $0.305^{+0.027}_{-0.031}$ & $0.331^{+0.021}_{-0.020}$ \\
$100\theta_{\mathrm s}$ & $1.04208\pm0.00054$ & $1.04212^{+0.00054}_{-0.00055}$ & $1.04205\pm0.00054$ & $1.04214^{+0.00058}_{-0.00055}$ & $1.04180\pm0.00058$ & $1.04186\pm0.00059$ \\
\hline
\end{tabular}
\caption{Marginalized constraints on cosmological parameters for the IDE model. $A$ and $B$ are the datasets defined in the main text, while the subscripts ${\rm no\text{-}SN}$ and ${\rm no\text{-}BAO}$ denote the corresponding dataset combinations obtained by removing the PantheonPlus SN and DESI BAO data, respectively.}
\label{tab:app3}
\end{table*}

Finally, we test whether the discrepancy between the full and compressed CMB analyses is driven by the particular combination of low-redshift datasets adopted in the main analysis. We repeat the comparison after removing either the SN or the BAO likelihood. The resulting marginalized constraints for IDE are reported in Table~\ref{tab:app3}.

The disagreement between the full and compressed CMB likelihoods persists when either low-redshift probe is removed. In some dataset combinations, the one-dimensional marginalized constraints reported in Table~\ref{tab:app3} may appear numerically comparable for several parameters. However, this apparent agreement does not extend to the full multidimensional posterior. The corresponding two-dimensional distributions can exhibit substantially different orientations and degeneracy directions, showing that the correlations among cosmological parameters are significantly altered by the CMB compression even when the individual marginalized intervals remain similar.

This behavior is illustrated in Figure~\ref{fig:app3}, where we compare the $A_{\rm no\text{-}BAO}$ and $B_{\rm no\text{-}BAO}$ combinations, obtained by removing the DESI BAO likelihood. Although some of the marginalized constraints are relatively close, the posterior contours in several parameter planes reveal a substantial change in the degeneracy directions. The magnitude of the discrepancy and the parameters most strongly affected depend on the dataset combination, as expected from the different degeneracies introduced by BAO and SN.

We have verified that the same qualitative behavior holds for the other cosmological models in which the compression fails in the main analysis.

\newpage
\bibliography{compression}

\end{document}